# Single condensation droplet self-ejection from divergent structures with uniform wettability

Nicolò G. Di Novo[1,2,*], Alvise Bagolini[2,*], Nicola M. Pugno[1,3,*]

[1] Laboratory of Bioinspired, Bionic, Nano, Meta, Materials & Mechanics, Department of Civil, Environmental and Mechanical Engineering, University of Trento, Via Mesiano, 77, 38123 Trento, Italy
[2] Sensors and Devices Center, Bruno Kessler Foundation, Via Sommarive 18, 38123 Trento
[3] School of Engineering and Materials Science, Queen Mary University of London, Mile End Road, London E1 4NS, United Kingdom
[*] corresponding authors

## Abstract
Coalescence-jumping of condensation droplets is widely studied for anti-icing, condensation heat transfer, water harvesting and self-cleaning. Another phenomenon that is arousing interest for potential enhancements is the individual droplet self-ejection. However, whether it is possible from divergent structures without detachment from pinning sites remains unexplored. Here we investigate the self-ejection of individual droplets from divergent, uniformly hydrophobic structures. We designed, fabricated and tested arrays of nanostructured truncated microcones arranged in a square pattern. The dynamics of the single condensation droplet is revealed with high speed microscopy: it self-ejects after cycles of growth and self-propulsion between four cones. Adopting the conical pore for simplicity, we modelled the slow iso-pressure growth phases and the surface energy release-driven rapid transients enabled once a dynamic configuration is reached. In addition to easier fabrication, microcones with uniform wettability have the potential to allow self-ejection of almost all the droplets with a precise size while maintaining mechanical resistance and thus promising great improvements in a plethora of applications.

## Introduction

Coalescence-induced condensation droplet jumping (also called coalescence-jumping) is a fascinating phenomenon observable on natural and artificial surfaces with a certain hydrophobicity and has been highly studied experimentally[1–4] and theoretically[5–9] in the last decade. On a sufficiently hydrophobic surface under condensation conditions, water droplets nucleate, grow, coalesce and eventually jump, leaving free space where the cycle begins again[10]. During coalescence the excess surface energy is transformed into kinetic energy of translation and oscillation[9], net of adhesive and viscous losses[5]. The detachment occurs on hydrophobic micro and/or nanostructured surfaces for their minimal adhesion but with less than 6% efficiency[11]. While for some surfaces of plants[12,13] and insects[14], coalescence-jumping contributes to self-cleaning from pathogens and inert particles, in the academic and industrial fields it is studied for several applications. For example, for dew harvesting from the atmosphere, coalescence jumping is an alternative[15] to grooves-induced distant coalescence[16,17] to enhance drainage. It also enhances heat transfer by condensation thanks to the continuous droplet shedding and re-nucleation of small droplets.[18,19] For these reasons, in frosting conditions, it also provides a passive anti-frost effect.[20,21] At negative temperatures, frost is often preceded by condensation.[22] Once few supercooled droplets freeze spontaneously (by homogeneous or heterogeneous nucleation[23–28]), frost percolates governed by the ice-bridging mechanism: the nearby liquid droplets evaporate and desublimate on the frozen one forming an ice bridge growing towards them.[20,22,26,29–31] The diameters and distances of the drops determine the success of the ice-bridging.[22,31–33] Coalescence-jumping slows down frost propagation because it inhibits successful ice-bridging.[21,34,35]
In recent years, the range of droplet sizes and environmental conditions that allow for coalescence-jumping has been expanded by new manufacturing ideas and techniques.[1,36,37] To avoid surface flooding with consequent loss of superhydrophobicity[38] and jumping ability[39], micro and nanostructured surfaces have been studied capable of spontaneously directing single droplets grown between the structures towards the apexes of the same and promote coalescence-jumping.[19,37,40,41]

A new topic of increasing interest is the single droplet jumping (here self-ejection) which broadens the knowledge in wettability and could potentially improve the mentioned applications and others. Individual droplets that grow confined in hydrophobic microstructures self-eject at a certain critical volume. The trigger can be an abrupt change of shape given either by the detachment of the drop from the bottom of the structures and by an enlargement in the upper part (micro-mesh[42] and rectangular grooves[43]), or the detachment from a strong pinning site (biphilic V-grooves[43]). Similarly, a rapid and stopped self-propulsion of single micro droplets was observed for vertical pillars[41], V-grooves[3,44], irregular cavities[40] and on conical threads[45], all microstructures having uniform wettability of the nanostructures. One interpretation[40,41,43,44] identifies the motion trigger with the overcoming of the retentive forces by the pressure force due to the Laplace pressure difference established, before detachment, inside the drop. On the other hand, it had already been observed that in quasi-static conditions (condensation) the drop grows with uniform internal pressure, albeit variable over time.[42] In support of this, Baratian et al.[46] first identified the system of the lone external forces acting on a static drop confined in a groove and proved, for a contact angle hysteresis-free surface, that it is null when the internal pressure is uniform. With the aid of simulations, both in the case with[47] and without[48,49] hysteresis, it was shown that a droplet in a groove moves spontaneously towards the iso-pressure configuration, coincident with the equilibrium one. In other words, before any self-propulsion or self-ejection, the droplet is static (mechanical equilibrium) and must respect the iso-pressure condition.

How to reconcile these experimental and theoretical evidences? Moreover, except for cases of abrupt changes in shape (depinning) with the clear establishment of a net pressure gradient, what forces trigger condensation droplets motion on uniform surfaces and when?

We make our contribution by studying a sufficiently general case that can be modelled analytically: a condensation droplet in a pore with conicity β and uniformly hydrophobic with certain dynamic contact angles ($\vartheta_a$ =157° and $\vartheta_r$ =145°, the advancing and receding angles of the tested microstructures walls, respectively). We first show that, regardless of the initial configuration, the droplet can accommodate the increasing volume by varying the contact angles and the position of the menisci while maintaining uniform internal pressure, which guarantees mechanical equilibrium, until it reaches a dynamic configuration. This configuration enables movement toward the opening, and as the surface energy ($E_{surf}$) of the drop decreases in that direction, a propulsive surface force ($F_{surf}$) sets the drop in motion. In other words, the energy of molecules no longer at the interface is converted into kinetic energy. Accounting for the opposing capillary (an external force exerted by the solid) and viscous forces, we numerically solve the rapid motion of micro droplets considering various dynamic angles and conicity of the pore: the drops accelerate and stop at a certain distance which depends on the three parameters. Then, the cycle repeats with the overall effect of a stick-and-slip movement towards the aperture. This analysis provides a first force-based explanation of the jerky motion of droplets growing in divergent structures and arrested self-propulsion in general. We infer that the release of surface energy is an irreducible driving force while the Laplace pressure gradients induced by abrupt shape changes eventually add to it in particular cases.

After that, the new aspect from a phenomenological point of view, as well as the main aim of the present study, is to investigate the not yet explored possibility of self-ejection of condensation droplets from diverging microstructures with uniform wettability. A certain tapering can be advantageous for various applications as it could reduce the percentage of droplets that grow on the apexes of the microstructures and therefore cannot exhibit self-ejection. Indeed, such droplets are limited to coalescence-jumping, a less efficient phenomenon which does not allow precise control of the jumping droplet size.

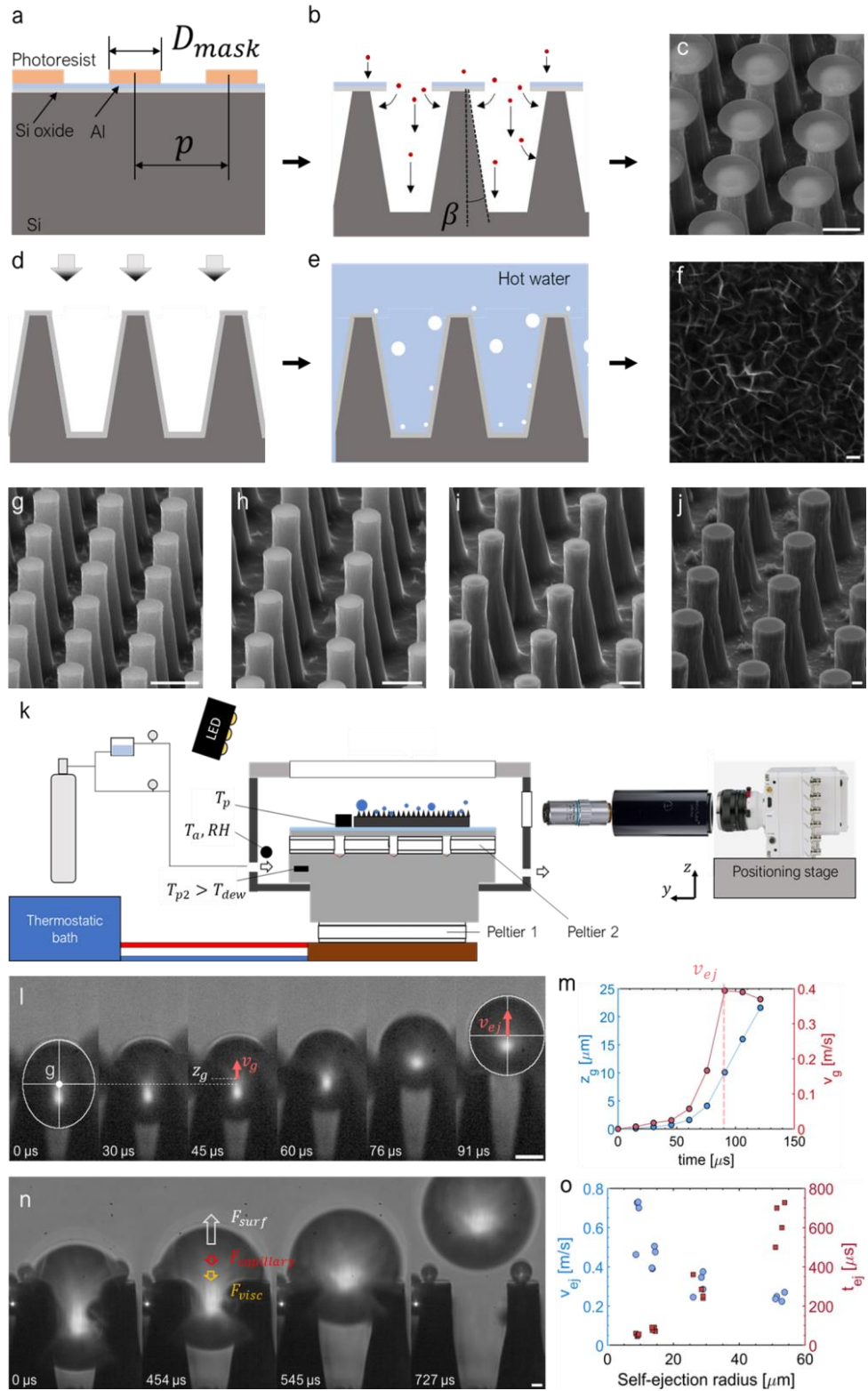

*Fig. 1. Fabrication steps and experiments. **a**) photolithography, **b**) tapered Reactive Ion Etching (t-RIE), **c**) Scanning electron microscope (SEM) image of Surface 15x20 after t-RIE, **d**) e-beam evaporation of Al, **e**) HWT and **f**) SEM image of NanoAl. SEM images of truncated microcones all tilted 60° respect to a plane ⊥ to the electron beam: **g**) Surface 10x13, **h**) 15x20, **i**) 30x40, and **j**) 60x80. **K**) Scheme of the experimental setup. **l**) Image analysis of a droplet self-ejecting from the nanostructured microcones of Surface 15x20 captured at 66 000 fps and **m**) the relative evolution of position and velocity of the centre of mass (g) assumed to be the centre of a fitted ellipse. **n**) Self-ejection event on the Surface_60x80 captured at 11000 fps (see Supplementary video 6). **o**) Scatter plot of the experimental self-ejection velocity ($v_{ej}$) and transient times ($t_{ej}$) measured for the four surfaces. Scale bars are 10 μm apart from the one of **f**) that is 100 nm.*

Furthermore, the absence of strong pinning sites (e.g., biphilic grooves) would greatly benefit the manufacturing. Considering a finite-length conical micropore, we model the droplet that reaches the edge of the aperture, after cycles of slow growth and self-propulsions. After another growth phase, the droplet reaches one of a range of dynamic configurations, it accelerates thanks to $F_{surf}$ and rapidly *self-ejects* with a velocity dependent on the droplet size, higher for smaller droplets.

We designed a type of microstructure that allows the observation of growth and self-propulsion of condensation droplets to verify the self-ejection predictions: arrays of solid, truncated microcones arranged in a square pattern. The choice lies in 4 factors: i) the drop between four conical pillars has axial symmetry and is similar to the modelled one, ii) the drop can be observed from the side, against the light, to capture the entire dynamics, iii) we avoid depressurisation effects of the air under the drop during movement (not modelled) which could be of some importance in the case of the pore and, iv) microcones have a small fraction of the top surface area and therefore we expect almost all the drops to nucleate between the cones and self-eject at a precise volume. We fabricated four kinds of truncated microcones having a tapering as similar as possible but different sizes by tapered reactive ion etching (t-RIE) of silicon, then covered by nanostructured aluminium (obtained by hot water treatment, HWT) and made hydrophobic (silanization). By testing the surfaces in condensation conditions, we captured the growth, stopped self-propulsion and self-ejection transients of single droplets with a high frame rate camera coupled with a microscope. The radius range of the droplets self-ejected from the four arrays is 9÷53 µm, with a correspondent self-ejection velocity ($v_{ej,exp}$) and transient time of 0.25÷0.65 m/s and 46÷620 µs, respectively, this being the first report of single drop movements with such spatial and temporal resolution. The pore model predictions are in good qualitative and quantitative agreement with the experiments. A critical examination of the differences is reported. For faster estimates of the self-ejection velocity and its dependence on the various parameters involved, we also develop an energetic model for a droplet between four solid truncated microcones.

Our study on divergent structures with uniform wettability describes with a novel approach the mechanisms underlying stopped self-propulsion and self-ejection, driven by the release of surface energy once a dynamic configuration is achieved by slow growth. Any Laplace pressure gradient established in the droplet after an abrupt shape change is only added to this basic mechanism but is not strictly necessary. Then, we demonstrate experimentally for the first time the self-ejection of single droplet from arrays of truncated microcones without pinning sites, a great advantage for large scale fabrication. Moreover, we deduce that ideal microcones would allow the self-ejection of all the droplets with a precise, designed size. This new way of designing and structuring expands the capabilities of superhydrophobic surfaces: in addition to coalescence jumping, it allows single drop self-ejection with precise size control and high energy conversion efficiency during jumping, promising important improvements in the various energy applications mentioned.

## Results

### Surfaces fabrication and characterization

*Table 1. Design ($D_{mask}$, $p$) parameters and final geometry ($\bar{\beta}, d_h$ and $l$) of the fabricated and tested surfaces.*

| Surface name | $D_{mask}$ [µm] | Pitch $p$ [µm] | Tapering $\bar{\beta}$ [°] | Head diameter $d_h$ [µm] | Height [µm] |
|---|---|---|---|---|---|
| 10x13 | 10 | 13 | 5.8 ± 0.7 | 5 | 23.3 |
| 15x20 | 15 | 20 | 5.7 ± 1 | 7 | 34.9 |
| 30x40 | 30 | 40 | 5.8 ± 0.6 | 12.5 | 64.3 |
| 60x80 | 60 | 80 | 5.5 ± 0.2 | 31.5 | 103.7 |

To study the growth, self-propulsion and self-ejection in divergent structures with uniform wettability we fabricated four arrays of truncated microcones arranged in a square pattern by photolithography on silicon with a mask layout composed by circles of diameter $D_{mask}$ and pitch $p$ (Table 1) which are identified with Surface_ $D_{mask}$ x$p$. The mask pattern was transferred to silicon using room-temperature t-RIE *(see*

Fig. 1.a-c and Methods and Supplementary information for fabrication details). The combined effect of anisotropic and isotropic etching results in pillars with a tapering ($\beta$), head diameter ($d_h$), height ($l$) and smoothness depending on the t-RIE process parameters, the area fraction free from the hard mask ($\varphi$) and the etching time ($t_e$). With the aim of fabricating four surfaces with the same $\beta$, $\varphi$ and aspect ratio $l/D$ but different size, we explored the effects of $\varphi$ and $t_e$ on $\beta$ and the etch rate by processing various Si wafers with a lithography mask consisting of 1 cm² light-exposed areas patterned with combinations of $D_{mask}$ and $p$. By analysing the parameters trends we selected four structures (Table 1 and Figure 1.g-j) realisable with similar geometry and the highest possible surface smoothness. We fabricated them on larger areas (2 cm x 10 cm) and cleaved in 2 cm x 2 cm samples. By evaporating pure Aluminium on the cleaved samples, followed by HWT, we obtained cones uniformly covered by Aluminium hydroxide nano flakes (NanoAl) (Figure 1.d-f). NanoAl was then rendered highly hydrophobic by conformal fluorosilane deposition (see Methods). We characterized the wettability of NanoAl by replicating the HWT and silanization procedures on flat silicon samples covered with evaporated Al. With the macro-droplet method $\vartheta_a$=166±1° and $\vartheta_r$=123±7° while the contact angles obtained with the micro-droplet method (the ones used in the modelling) are $\vartheta_a$=157±1 and $\vartheta_r$=145±6° (see Methods). We placed the surfaces on a cold plate inside a chamber with controlled humidity and observed the dynamics of condensation droplets with a high frame rate camera coupled with a microscope (see Figure 1.k and Methods).

### Single droplet growth between truncated microcones, self-propulsion and self-ejection

The condensation droplets nucleate on random sites that can be either the lateral and bottom walls or truncated cones heads. Apart from the last case, the droplet grows and touches the inner walls of the four cones and, after a certain time, settles in the axial symmetric position. In this phase, the droplet moves toward the aperture by alternating a slow growth (via condensation) and fast *self-propulsions* when a dynamic configuration is reached (Suppl. Video 1-2). After a time which depends on the unit cell size (in our cases in the order of minutes), the droplet arrives at the top edges, slowly grows to another dynamic configuration (Suppl. Video 2) and rapidly *self-ejects* (Figure 1.l and n and Suppl. Video 3-6). Depending on its volume (radius of an equivalent sphere in the range 9÷53 $\mu$m), the droplet accelerates to ~0.25÷0.65 m/s (*self-ejection velocity*) in ~40÷700 μs, respectively, and detaches from the structures (Figure 1.l-o). Four self-ejection events are reported for each surface. Fig. 1.l shows how we acquired the evolution of the centre of mass position and velocity during self-ejection transients (see Methods). By analysing the Supplementary Videos 1-6 (side-view), 7 (top-view) and 8-9 (side-view not perpendicular to the cleavage line, see Figure S1), the drop resembles a spheroid that touches the four walls of the cones and partially touches the truncated heads. The contact areas are therefore pseudo-elliptical non-flat surfaces which evolve during motion, a very tricky case to deal with analytically. The modelling of the drop in a hydrophobic conical pore, on the other hand, makes the problem analytically treatable and, however far it may be quantitatively from the real case, preserves the qualitative aspects, as we shall explain.

### The system of the external forces

With the aim of describing the quasi-static growth and simulating the rapid motions of the droplet considered as a particle, we first analyse the forces involved. Let us consider the droplet suspended in the conical pore during growth (Figure 2.a) and verify that the Laplace pressure and contact line forces acting on the "*top meniscus system*" (Figure 2.b) are equal and opposite. The droplet is micrometric (radius < 100 μm) and gravitational effects can be safely neglected. Being the Laplace pressure $\Delta P_t \equiv P_{int} - P_{ext} = -[2\sigma_{lv}\cos(\vartheta_t + \beta)]/r_t$, the upward force $F^{\uparrow}_{Laplace}$ is $\Delta P_t \cdot A$, where $A = \pi r_t^2$ is the spherical cap area projected on a plane ⊥ to z with contact radius $r_t$, $\sigma_{lv}$ is the liquid-vapour surface tension, $\beta$ is the half-aperture, and $\vartheta_t$ the apparent contact angle on the wall. Thus, $F^{\uparrow}_{Laplace,t} =$

$-2\pi\, \sigma_{lv}\, r_t\, cos\, (\vartheta_t + \beta)$ , a force exerted by the droplet bulk on the meniscus system along the positive z axis. The force generated by the reaction of the solid to the surface tension acting on the circular contact line[50], projected along positive z, is $F_{\sigma,t}^{\uparrow} = -2\pi\, \sigma_{lv}\, r_t\, cos(\pi - \vartheta_t - \beta)$, equal in modulus to $F_{Laplace,t}^{\uparrow}$ but opposite. The proof for the *bottom* meniscus is analogous. This result confirms the force balance of the generic meniscus and is, in fact, true by definition. Therefore, differently from other studies[43], we do not put both of these terms into the system of external forces acting on the "*droplet*" system because they would cancel out.

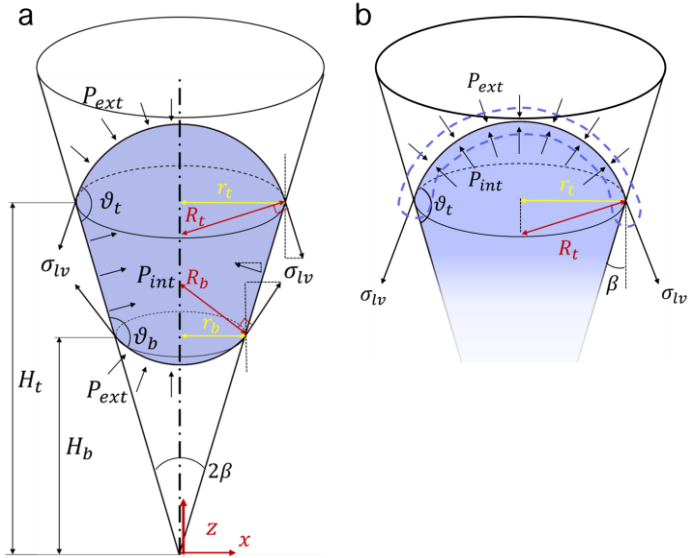

*Figure 2. **a**) Scheme of a droplet suspended in a hydrophobic conical pore showing half of the forces acting on the droplet system in the xz plane: surface tension on the contact lines and pressures on the contact area and caps . **b**) Forces acting on the meniscus system, highlighted with the blue dotted line.*

Before proving that a droplet suspended in a conical pore with hysteresis satisfies the mechanical equilibrium when the internal pressure is uniform, we report other studies that identified the iso-pressure as an equilibrium condition for similar geometries. Molecular dynamics simulations[51] of a droplet growing in a zero-hysteresis wedge identified the truncated sphere (constant curvature) as an equilibrium configuration, thus with uniform internal pressure. Baratian et al.[46] first described analytically that, on such a droplet, the system of external forces acts with null resultant. A drop placed away from this configuration experiences an internal pressure gradient (with a quasi-linear profile) and correspondingly a net force that brings it to the equilibrium configuration (iso-pressure), as described analytically[48] and with Lattice-Boltzmann simulations[49]. On the other hand, as highlighted in experiments and modelling[47,52–54], the role of hysteresis is essential for a more refined description of equilibrium and motion of droplets in wedges.

Now, regarding the "*droplet system*" suspended in a conical pore with hysteresis, the contact angles $\vartheta_t$ and $\vartheta_b$ vary in the range $[\vartheta_r, \vartheta_a]$, and by imposing the curvature radii $R_t$ (Eq. 1) and $R_b$ (Eq. 2) to be equal, we obtain Eq. 3, the relationship to be satisfied by the five parameters $H_t$, $H_b$, $\vartheta_t$, $\vartheta_b$ and $\beta$ in order to have a uniform internal pressure. In Eq.3 we define the *shape ratio* $\lambda \equiv H_t/H_b$ between the heights of the contact lines. The external forces are: the reaction that the solid opposes to the surface tension acting on the contact lines of both menisci, $F_{t,\sigma}^{\uparrow}$ and $F_{b,\sigma}^{\uparrow}$ (Eqs. 4-5), the reaction of the solid to the droplet internal pressure acting on the contact area, $F_p^{\uparrow} = (P_{ext} + \Delta P_{Laplace}) A_{truncated\ cone} sin\, \beta$ , and the atmospheric pressure force on caps ($F_{cap}$), all projected along the positive verse of the z symmetry axis. However, note that the constant atmospheric pressure $P_{ext}$ acting on both caps and contact area (a close surface) generates a null net force based on the Gauss's theorem. Thus, we do not consider $F_{caps}$ and directly use Eq. 6 for $F_p^\uparrow$ where $\Delta P_{Laplace} = \frac{2\,\sigma_{lv}}{R_t}$, uniform in the droplet.

$$R_t = -\frac{H_t \tan\beta}{\cos(\vartheta_t + \beta)} \tag{1}$$

$$R_b = -\frac{H_b \tan\beta}{\cos(\vartheta_b - \beta)} \tag{2}$$

$$\lambda \equiv \frac{H_t}{H_b} = \frac{\cos(\vartheta_t + \beta)}{\cos(\vartheta_b - \beta)} \tag{3}$$

$$F_{t,\sigma}^\uparrow = -2\pi\,\sigma_{lv}\,H_t \tan\beta\,\cos(\pi - \vartheta_t - \beta) \tag{4}$$

$$F_{b,\sigma}^\uparrow = 2\pi\,\sigma_{lv}\,H_b \tan\beta\,\cos(\pi - \vartheta_b + \beta) \tag{5}$$

$$F_p^\uparrow = \frac{\pi(r_t + r_b)(H_t - H_b)}{\cos\beta} \cdot \frac{2\,\sigma_{lv}}{R_t} \sin\beta \tag{6}$$

The resultant capillary force acting on the droplet, $F_{capillary}^\uparrow = F_{t,\sigma}^\uparrow + F_{b,\sigma}^\uparrow + F_p^\uparrow$, is identically null for every $H_t, H_b, \vartheta_t, \vartheta_b$ and $\beta$ that satisfy Eq. 3 (see S1, Supplementary Information). In other words, given a droplet of a certain volume in the pore, the configurations such that the internal pressure is uniform are also equilibrium configurations for the external force system. During the slow growth by condensation, that is a quasi-static process, and before dynamic conditions, it is reasonable to assume that the incoming volume redistributes in the possible equilibrium configurations (Eq.3) by varying the contact line heights and contact angles. Of course, the surface energy required to enlarge the interfaces comes from the condensation volume entering the system.

### Growth phases in a conical pore

The droplet nucleates somewhere on the walls, it swells and increases the contact area until it equilibrates axis symmetrically in the pore (or similarly between the 4 cones as shown in Video S7). Describing analytically what takes place during the process of symmetric settling is not an easy task and the present scope. However, after this settling we can say that $\lambda \in [\lambda_{min,growth}, \lambda_{max}]$ with the extremal shape ratios defined in the Eqs.7-8 and marked with a black and a red dot, respectively, in Figure 3.A. In Figure 3.A, we represent the relation between $\vartheta_b$ and $\vartheta_t$ (from Eq. 3) for the various $\lambda$ allowed. As the volume increases, assuming that the drop passes through iso-pressure configurations, $\vartheta_b, \vartheta_t, H_b$ and $H_t$ evolve following precise paths. We identified two subsequent growth phases: phase 1 which consists of two sub-cases both of which lead to $\vartheta_t = \vartheta_a$, and phase 2 that ends with $\vartheta_t = \vartheta_a$ and $\vartheta_b = \vartheta_r$, a dynamic configuration with the shape ratio $\lambda_{max}$.

$$\lambda_{min,growth} = \frac{\cos(\vartheta_r + \beta)}{\cos(\vartheta_a - \beta)} \tag{7}$$

$$\lambda_{max} = \frac{\cos(\vartheta_a + \beta)}{\cos(\vartheta_r - \beta)} \tag{8}$$

*Growth phase 1*

In phase 1, for any particular couple of $\vartheta_b, \vartheta_t \in (\vartheta_r, \vartheta_a)$, as the volume increases, the contact angles can only increase constrained to a black curve as reported in figure 3a. Depending on the initial $\lambda$ either $\vartheta_b$ or $\vartheta_t$ reaches $\vartheta_a$ first. For the particular $\lambda^* = \frac{\cos(\vartheta_a + \beta)}{\cos(\vartheta_a - \beta)}$, they reach $\vartheta_a$ contemporarily (violet dot) at the end of phase 1.

*Sub-case 1*: $\lambda \in [\lambda^*, \lambda_{max}]$. $\vartheta_t$ reaches $\vartheta_a$ first (green dots in Figure 3.a) and the droplets enter directly into *phase 2*.

*Sub-case 2*: $\lambda \in [\lambda_{min,growth}, \lambda^*)$. The steps are: $\vartheta_b$ reaches $\vartheta_a$ first (blue dots in Figure 3.a), the bottom meniscus advances towards the tip of the pore, $H_b$ decreases and thus $\lambda$ increases towards $\lambda^*$. In parallel, $\vartheta_t$ increases towards $\vartheta_a$. When $\vartheta_t = \vartheta_a$ both menisci are in the advancement condition but, in the subsequent growth only $H_t$ can increase, otherwise $R_b$ could not be equalled by $R_t$. Again, the system enters phase 2. For the limit initial shape ratio $\lambda_{min,growth}$, the only difference is the absence of the first step.

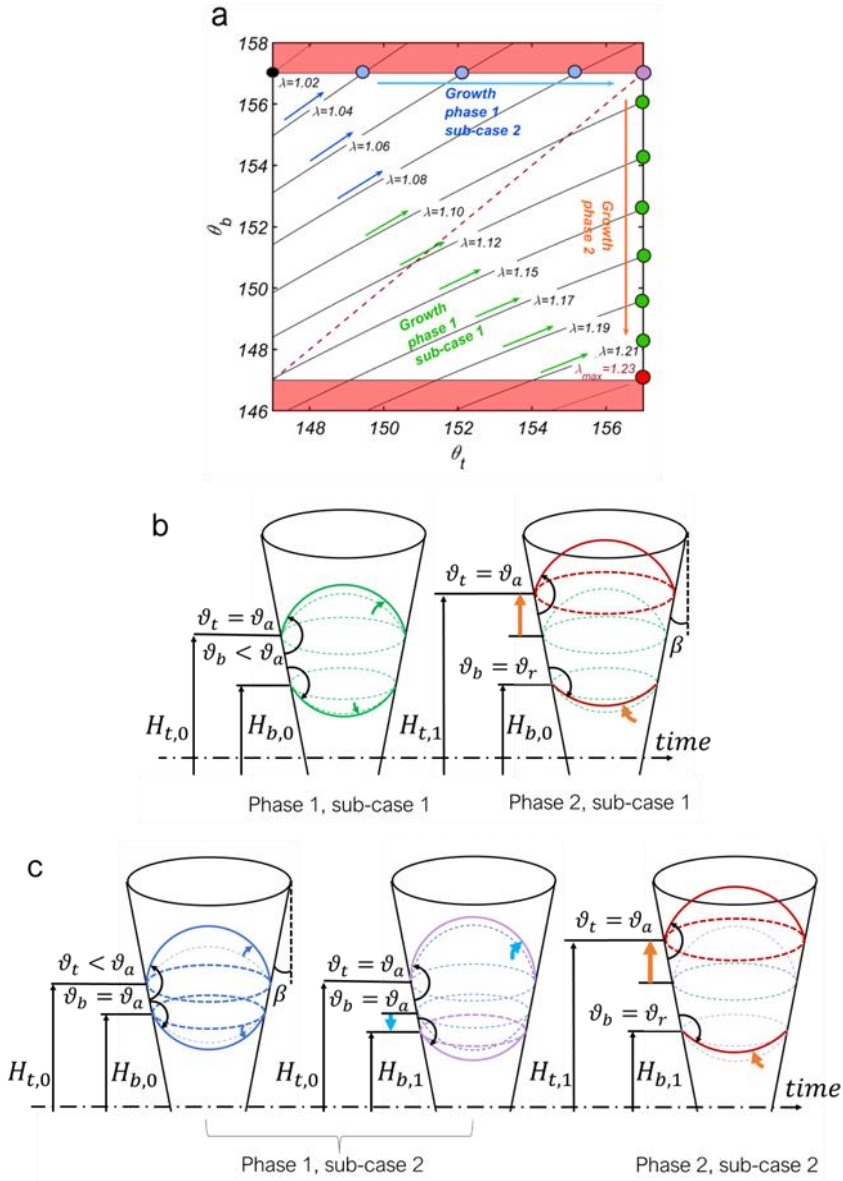

Figure 3. **a**) Diagram of the growth phases 1 (with the two sub-cases) and 2 in terms of $\vartheta_b$ and $\vartheta_t$ for various initial shape ratios $\lambda \in [\lambda_{min,growth}, \lambda_{max}]$ where these two extremal values are depicted as a black and a red dot, respectively, for $\beta = 6°$. Any point on a black line, in the contact angles range $[\vartheta_r, \vartheta_a] = [147°, 157°]$, represents a possible initial configuration (shadowed regions are not allowed). As the volume increases during phase 1, the system can either reach a blue, violet or green dot following its black line. Then, during phase 2, $\lambda$ increases to $\lambda_{max}$, the dynamic configuration (red dot), regardless of the initial configuration, and the droplet self-propels. The red dotted line represents $\vartheta_b = \vartheta_t$. In **b**) and **c**) illustrations are reported where caps depicted with dotted lines indicate configurations before the ones with solid lines, and colours are referred to the ones of arrows and dots in **a**).

*Growth phase 2*

Starting from $\lambda^*$ or any other green dot, $H_t$ increases maintaining $\vartheta_t = \vartheta_a$ while $H_b$ is fixed and $\vartheta_b$ decreases to $\vartheta_r$; in other words, $\lambda$ increases towards $\lambda_{max}$. At the end, regardless of the initial $\lambda$, all droplets reach $\lambda_{max}$. Growth *phase 2* is clearly observed in Supplementary Videos 1-2, where, before self-propulsion, only the *top* meniscus moves. The shape ratio $\lambda_{max}$ corresponds to the *dynamic configuration* of self-propulsion. Once reached, both menisci can move and the release of $E_{surf}$ begins. The theoretical $\lambda_{max}$=1.31±0.13 gives an estimate of the experimental one ($\lambda_{max,exp}$= 1.19±0.02), calculated by analysing the droplet shape before self-propulsion with the use of the pore model assumptions (S2, Supplementary information).

## Self-propulsion transient in a conical pore

We here show that self-propulsion causes a small displacement, a fraction of the droplet size, without detachment, as observed in our experiments but also in the ones of droplets in wedges[3,44]. In this section we develop the theoretical framework for the resolution of the droplet motion which is then taken up in the self-ejection section in which we directly compare the theoretical and experimental motion.

During the growth by condensation in a confined space, the droplet surface energy, $E_{surf}$, is stored in the geometrical configuration dependent on the contact angle hysteresis and compatible with the iso-pressure condition. Once $\lambda_{max}$ is attained, the motion is enabled and the system evolves towards configurations with lower $E_{surf}$. We consider the volume to be constant in these transients (both self-propulsion and self-ejection) as the motion occurs in tens to hundreds of $\mu s$ and the condensation volume is negligible. Also, we assume the liquid to be incompressible. The droplet volume in such a pore is $V = V_{cap,t} + V_{cap,b} + V_{tc}$, Eq.9, expressed as the sum of the top and bottom spherical caps, and of the truncated cone volumes, respectively (Eqs. S 3.1-3). The self-propulsion volume $V^*$ of a particular simulated droplet can be calculated with Eq. 9 by choosing $\vartheta_b = \vartheta_r$, $\vartheta_t = \vartheta_a$ and an initial $H_b$, called $H_{b,0}$, depending on the condensation nucleus location in the pore and an initial $H_t$, called $H_{t,0}$, which is $\lambda_{max}H_{b,0}$. During sliding towards the aperture, $H_b$ and $H_t$ vary but the contact angles and the volume $V^*$ remain constant. Thus, by imposing $V^* = V$ (Eq. 10), with $V$ of the general Eq. 9, $H_t$ depends on $H_b$ during all the movement (Eq. S3.4). For each examined droplet, every quantity expressed below is only a function of $H_b$ and its time derivatives. $E_{surf}$ is expressed in Eq. 11 (explicit form in S3.4) where $\vartheta_{eq} = \cos^{-1}[(\cos\vartheta_a + \cos\vartheta_r)/2]$ is an estimate of the equilibrium contact angle. Under these hypothesis, $E_{surf}$ is the potential of a driving surface force $F_{surf}$ (Eq. 12)[55–58], positive as $H_b$ increases for the surface considered here.

$$E_{surf} = \sigma_{lv}\left(A_{cap,t} + A_{cap,b} - A_{truncated\ cone}\cos\vartheta_{eq}\right) \qquad (11)$$

$$F_{surf}^\uparrow = -\frac{dE_{surf}}{dH_b} \qquad (12)$$

At the same time, the external force system, $F_{capillary}^\uparrow(H_b)$, acts on the droplet. $F_{b,\sigma}^\uparrow$ is the same of Eq. 5 while for $F_{t,\sigma}^\uparrow$ we substitute $H_t$ of Eq. 4 with Eq. S 3.5. As the droplet moves away from the iso-pressure configuration, the two curvatures evolve differently and thus there is an internal pressure gradient. We modify Eq. 6 to Eq. 13 by assuming a linear pressure profile $P(z) = p_b + (\Delta p_t - \Delta p_b)(z - H_b)/(H_t - H_b)$, (Eq. 14), in similarity to what was found elsewhere for wedges[49]. The infinitesimal area of the truncated cone is $dA = 2\pi \tan(\beta)\, z\, dz/\cos\beta$ with $z \in [H_b, H_t]$. $\Delta p_t = 2\sigma_{lv}/R_t$, $\Delta p_b = 2\sigma_{lv}/R_b$ and $p_b = \Delta p_b + p_{ext}$ are the top and bottom Laplace pressure differences and the bottom pressure, respectively. Again, the force contribution of $p_{ext}$ on the truncated cone area balances with those on the two caps so, for calculation purposes, we place it equal to zero in Eq. 13 and do not consider $F_{caps}$. With substitutions and by solving the integral in $z$, also $F_p^\uparrow$ is a function of $H_b$ alone.

$$F_p^\uparrow = \int_A P(z)\sin\beta\, dA = \int_{H_b}^{H_t}\left[\Delta p_b + \frac{(\Delta p_t - \Delta p_b)(z-H_b)}{(z-H_b)}\right]2\pi(\beta)\, z\, dz \tag{13}$$

$F_{capillary}^\uparrow$ is negative as the droplet moves towards the aperture (opposing force) while the net force $F_{net}^\uparrow = F_{capillary}^\uparrow + F_{surf}^\uparrow$ (Eq. 15) is positive until $H_b = H_b^*$ (Figure 4.a): the droplet accelerates, decelerates and stops. The motion is therefore driven by a position-dependent force $F_{net}^\uparrow(H_b)$ acting on the droplet centre of mass $H_g$, which is also a function of the lone $H_b(t)$ (see S3, Supplementary Information). In addition, the walls oppose a viscous force $F_{visc} = \tau \cdot A$ to the droplet as it slides, with $\tau$ and $A$ being the shear stress and contact area, respectively. By assuming a Poiseuille flow in a tube of radius $r$ and for small $\beta$, the fluid velocity profile can be approximated as $v(x) = \dot{H}_g\left[1 - \left(\frac{x}{r}\right)^2\right]$, $\tau = \mu \frac{dv}{dx}\Big|_{x=r}$, $A \approx 2\pi r(H_t - H_b)$, and the viscous force[59] as

$$F_{visc} = 4\pi\mu\dot{H}_g(H_t - H_b) \tag{16}$$

where $\dot{H}_g$ is the velocity of the centre of mass and $\mu$ is the dynamic viscosity. Considering the drop as a particle accelerating under the effect of the described forces, we obtain Eq. 17, where $\rho$ is the water density and $\ddot{H}_g$ the acceleration of the droplet centre of mass. By substituting the expressions of $\dot{H}_g$ and $\ddot{H}_g$ as functions of $H_b$, $\dot{H}_b$ and $\ddot{H}_b$ into Eq. 17 and solving numerically in MATLAB, we obtained $H_b(t)$ and $\dot{H}_b(t)$ (details in S3, Supplementary information). The developed code detects particular events such as reaching a prescribed $H_b^{\#}$ or $\dot{H}_b = 0$ and stops the resolution. Then, by substituting the $H_b$ and $\dot{H}_b$ numerical values in $H_g$ and $\dot{H}_g$, we plotted the motion of the centre of mass (example in Figure 4.b).

$$\ddot{H}_g \rho V^* = F_{net}^\uparrow - F_{visc} \tag{17}$$

The droplet does not stop at $F_{net}^\uparrow = 0$ (identified by $H_b^*$ in Figure 4.a-b) but at a certain final $H_{b,fin}$. For self-propulsion, the resolution must be interrupted when the drop stops ($\dot{H}_b = 0$) because before any eventual acceleration in the opposite direction, the contact angles should be reconfigured (*bottom* in advancing and *top* in receding conditions) and the capillary force system rewritten. In the stop position, $F_{capillary}^\uparrow < 0$ because $F_{t,\sigma}^\uparrow$ is negative and greater in modulus than the positive $F_{b,\sigma}^\uparrow + F_p^\uparrow$ and thus the internal pressure is not uniform. In particular, the meniscus top has a larger pressure. Before the eventual motion towards the negative direction the contact lines are fixed and the contact angles rearrange to cancel the internal pressure difference. If $\lambda_{fin} = H_{t,fin}/H_{b,fin}$ does not allow an iso-pressure configuration, given the particular fixed volume $V^*$, the droplet reaches the apt dynamic configuration for travelling in the opposite direction, it accelerates, stops and so on. For the parameters $\vartheta_a, \vartheta_r$ and $\beta$ of interest in this study we found that the droplets can re-equilibrate in the stop position (identified by $H_{b,fin}$) and calculated which final $\vartheta_{t,fin}$ and $\vartheta_{b,fin}$ are attained by the menisci starting from $\vartheta_a$ and $\vartheta_r$, respectively (S3.4, Supporting Information). Figure S1 shows $H_{b,fin}/H_{b,0}$, $\vartheta_{t,fin}$ and $\vartheta_{b,fin}$ for various equivalent radii $R_{eq}$ (the radius of a spherical droplet with the same volume $V^*$), for both the viscous and non-viscous cases and fixed surface parameters; it is interesting to note that all three quantities are independent of droplet size for the non-viscous case while varying marginally in the viscous one. With multiple simulations we built a 3D map of $H_{b,fin}/H_{b,0}$ for various $\vartheta_a, \vartheta_r$ and $\beta$, considering the non-viscous case (Figure 4.c): as a guideline, we deduce that $H_{b,fin}/H_{b,0}$ increases with $\vartheta_a$ and with the contact angle hysteresis, while it decreases with $\beta$. In addition to those shown in Supplementary video S1-2, we captured and analysed four other self-propulsion events. All the videos were captured orthogonally with respect to the cleavage line (see Figure 4.d and Figure S1) from which the actual point of contact of the meniscus bottom with the walls cannot be seen; the ideal view would be at 45° but it does not allow good illumination of the meniscus bottom as the light is blocked by the cones behind it; we therefore measured $h_{b,0}$ and $h_{b,fin}$ (Figure 4.d) and calculated the experimental $H_{b,fin}/H_{b,0}=1.05\pm0.02$ (analogously to S2, Supplementary information), overestimated by ~8% by the viscous model that predicts $H_{b,fin}/H_{b,0}\sim 1.14$ for the droplet captured. Transient time of self-propulsion is not the main scope of the present article; however, we captured at 2000 fps the self-propulsion of a droplet with $R_{eq}$ similar to the simulated one (Figure 4.a) and it takes place between two frames, thus in less than 500 µs.

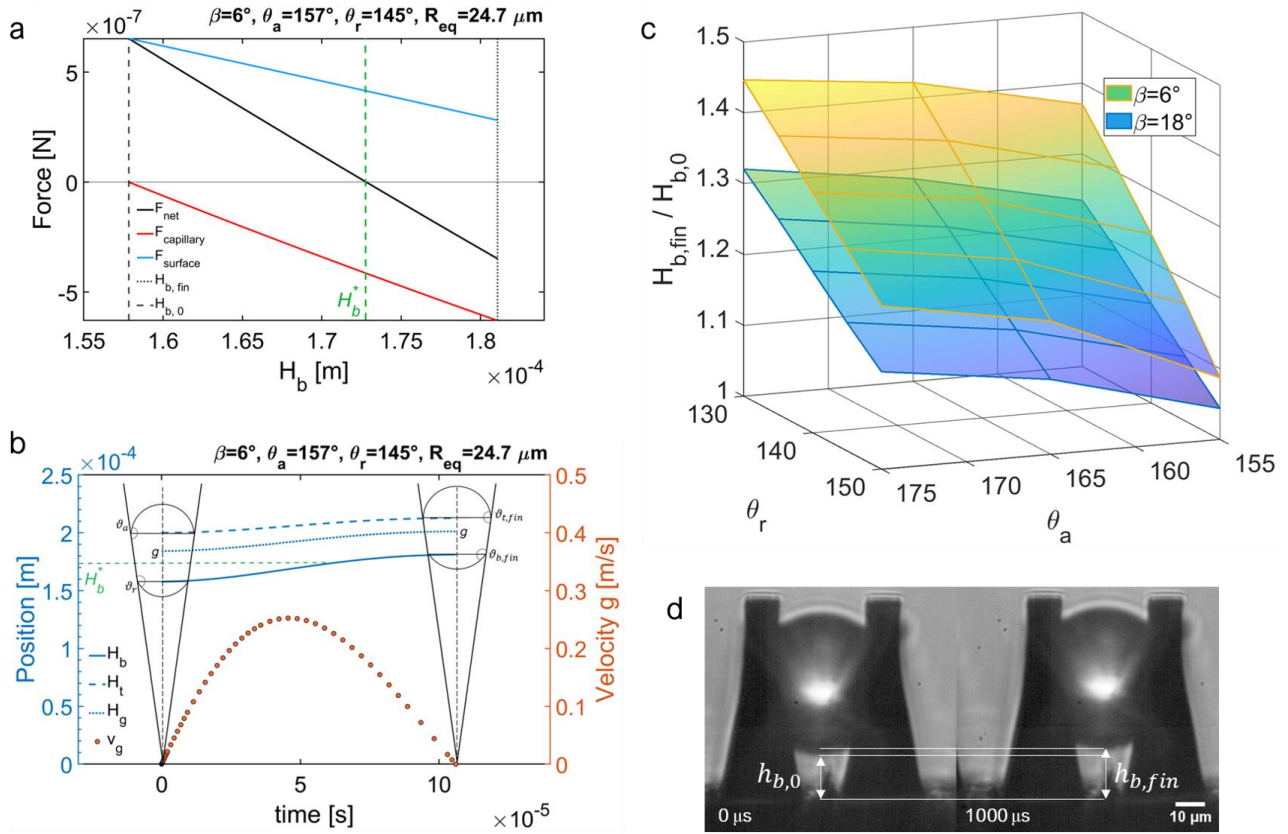

Figure 4. **a**) $F^{\uparrow}_{net}$, $F^{\uparrow}_{capillary}$, $F^{\uparrow}_{surf}$ as a function of $H_b$ for a droplet with equivalent radius $R_{eq}=24.7$ µm, $\vartheta_a=157°$, $\vartheta_r=145°$ and $\beta=6$. $H^*_b$ (green dotted line) corresponds to $F^{\uparrow}_{net}=0$. $H_{b,0}$ is the initial condition while its final value, $H_{b,fin}$, comes from the resolution of Eq.17 stopped at $\dot{H}_b=0$. **b**) Time evolution of the droplet considering viscous dissipations: position of $H_b$, $H_g$ and $H_t$ respect to the cone apex and the velocity of the centre of mass, $\dot{H}_g$. **c**) 3D map of $H_{b,fin}/H_{b,0}$ vs $\vartheta_a, \vartheta_r$ and $\beta$ for the non-viscous case. **d**) Example of droplet self-propulsion recorded at 1000 fps by testing a Surface_30x40 and illustration of the measurement of $h_b$.

## Self-ejection from a conical pore

The central theme of the present article is the droplet spontaneous self-ejection from arrays of micro truncated cones covered by uniformly hydrophobic nanostructures. Once nucleated and settled between four truncated cones, the droplet alternates between growth phases and self-propulsions until the *top* meniscus reaches the top edges ($H_{t,max}$) with a particular shape ratio ($\lambda_{ej}$), not unique but history-dependent. In the framework of the conical pore approximation, the subsequent increase in volume by condensation or acceleration does not imply the advancement of the *top* meniscus (fixed at $H_{t,max}$) but the increase of $\vartheta_t$ (towards $\vartheta_a$ measured with respect to the plane orthogonal to z). As the volume increases by condensation, the two menisci evolve to overcome the striction imposed by the edge until *bottom* reaches $\vartheta_r$ (the *self-ejection dynamic configuration*) and the droplet self-ejects by releasing $E_{surf}$. We employ again the conical pore approximation to describe the pre-self-ejection growth and the self-ejection transient. The differences between the real case and the model will be examined in the energetic model paragraph.

*Pre-self-ejection growth*

Again, we consider the growth as a quasi-static process with uniform pressure at each instant. $\lambda_{ej}$ can be between two extreme values to be identified among three possible cases. **Case 1**: The drop is exactly at the end of growth phase 2 which would result in self-propulsion if the droplet is not at the edge; therefore, when $H_t = H_{t,max}$, $\lambda_{ej,1} = \lambda_{max}$ (Figure 5.a). **Case 2**: the droplet has just finished a self-propulsion and stops exactly with $H_{t,fin} = H_{t,max}$; $\lambda_{ej,2} = \lambda_{fin}$, $\vartheta_b = \vartheta_{b,fin}$ and $\vartheta_t = \vartheta_{t,fin}$. **Case 3**: the droplet nucleated near the edge ($H_t$ is already equal to $H_{t,max}$) and only goes through growth phase 1 (sub-case 2) (Figure 5.b). The contact angles evolution of each case is described in detail here:

- **Case 1**. As the volume increases, $\vartheta_b$ first increases and then returns to $\vartheta_r$ once the striction is overcome (Figure 5.c, obtained from Eq. 3 with $H_t = H_{t,max}$ and $H_b = H_{t,max}/\lambda_{max}$). When $\vartheta_b = \vartheta_r$ again, $\vartheta_t = \hat{\vartheta}_{t,1} > \vartheta_a$ (contact angle in reference to the pore walls), the droplet has a certain volume $V_{ej,1}$ (calculated with Eq. 9 and the appropriate parameters) and the self-ejection transient begins.
- **Case 2**. Same development as in *Case 1* but with different initial contact angles and $\lambda_{ej}$. Once the striction is overcome and $\vartheta_b$ decreases to $\vartheta_r$ the droplet is enabled to move but $\hat{\vartheta}_{t,2} \neq \hat{\vartheta}_{t,1}$ and $V_{ej,2} \neq V_{ej,1}$. Therefore, as we shall show, the transient and ejection velocity is different from *Case 1* even if the surface parameters are the same.
- **Case 3**. The droplet goes through *growth phase 1* (*sub-case 2* without the limit $\vartheta_a$ for $\vartheta_t$) with an advancing *bottom* meniscus until $\vartheta_t = \pi - \beta$ which corresponds to the minimum $R_t$ imposed by the striction and to $\lambda_{ej,3} = \lambda_{min} = -1/\cos(\vartheta_a - \beta)$. Then, as the volume increases, $\lambda_{ej,3}$ is maintained, $\vartheta_b$ decreases to $\vartheta_r$ and $\vartheta_t$ increases to $\hat{\vartheta}_{t,3} \neq \hat{\vartheta}_{t,2} \neq \hat{\vartheta}_{t,1}$. Also $V_{ej,3} \neq V_{ej,2} \neq V_{ej,1}$. For $\beta, \vartheta_a$ and $\vartheta_r$ of our surfaces (Figure 5.c), $\lambda_{min} \leq \lambda_{fin}$.

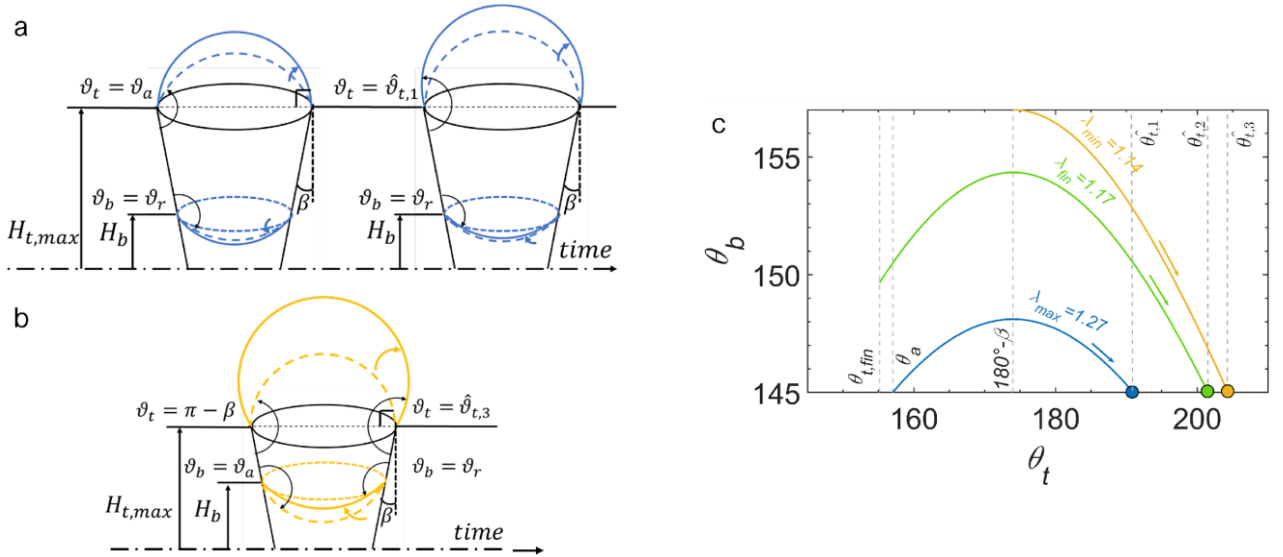

*Figure 5. Schemes of the pre-self-ejection growth in     Case 1 (**a**) and Case 3 (**b**). The caps depicted with dotted lines indicate earlier configurations than those with solid lines and colours are referred to **c**) where the contact angles evolution of the three cases is plotted for the case $\beta$=6°, $\vartheta_a$=157° and $\vartheta_a$=145°.*

The pre-self-ejection growth analysis and Figure 5.c reveal that, for any fixed $\beta, \vartheta_a, \vartheta_r$ and $H_{t,max}$, the ejection shape ratio ($\lambda_{ej}$), the ejection volume ($V_{ej}$) and the *top* contact angle at the beginning of motion ($\hat{\vartheta}_t$) depend on the initial bottom meniscus height, $H_{b,0}$, which determines the number of self-propulsions (zero, one, two or more) the droplet will make to reach the edge. Since $H_{b,0}$ is a random variable depending on the nucleation site, we will study the two extreme cases being $\lambda_{min} \leq \lambda_{ej} \leq \lambda_{max}$.

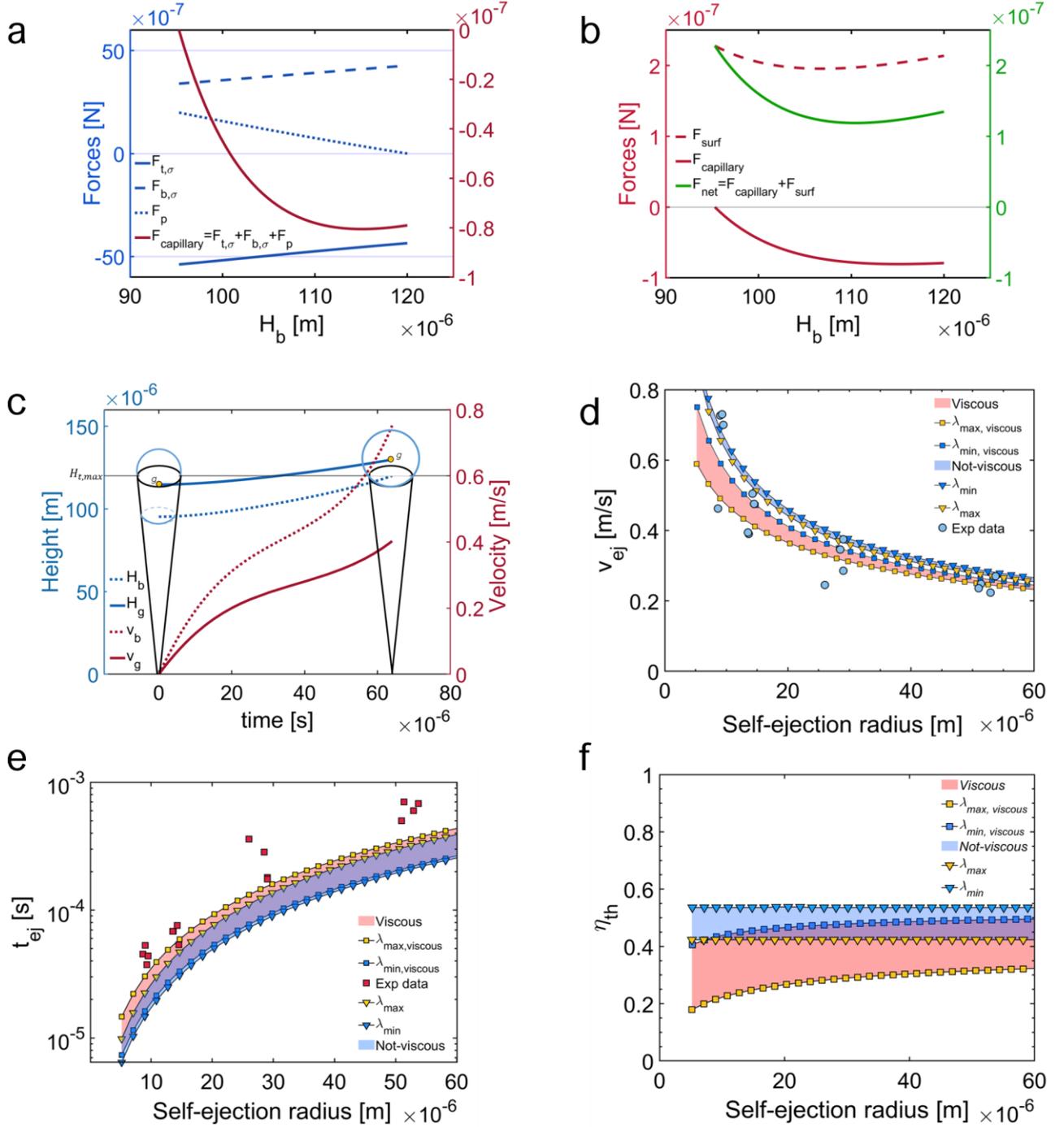

*Figure 6. Evolution of forces (**a-b**), $H_b(t), H_g(t)$ and their velocities (**c**) for a droplet with $R_{eq} = 14.7$ μm and $\lambda_{max}$. Comparison of the experimental data with **d**) the theoretical ranges of the final self-ejection velocity of the centre of mass and **e**) transient time for different equivalent radii for both the viscous and non-viscous cases and fixed surface parameters $\beta$=5.7°, $\vartheta_a$=157° and $\vartheta_a$=145°. **f**) Theoretical efficiency $\eta_{th}$ as the equivalent radii vary for both the viscous and non-viscous cases.*

*Self-ejection transient*

Once the drop has reached one of the possible *dynamic configurations* (a certain $\lambda_{ej}$ to which certain $\hat{\vartheta}_t$ and $V_{ej}$ correspond) and accelerates under the effect of $F_{net}^{\uparrow} - F_{visc}$, the displacement at constant volume occurs with an increase of $H_b$ at constant $\vartheta_b = \vartheta_r$ and with an increase of $\vartheta_t$ at constant $H_t = H_{t,max}$. As done for self-propulsion, $V_{ej} = V$ (Eq. 18), with $V$ of the general Eq. 9 and $V_{ej}$ a numerical value; being $\vartheta_b$ and $H_t$ constants, $\vartheta_t$ depends on $H_b$ during the motion (see S4, Supporting Information). It follows that all the other quantities in Eqns. 15-16 are also functions of $H_b(t)$ alone. For the experimental $\beta, \vartheta_a$ and $\vartheta_r$ and for various $H_{t,max}$, we solved numerically Eq. 17 (in the variables $H_b(t)$ and its time derivatives) for the two extreme shape ratios $\lambda_{min}$ and $\lambda_{max}$ and stopped the resolution when the condition $H_b = H_{t,max}$ is verified (the instant in which the droplet has a null contact area and detaches). Figures 6.a-b show the evolution of the forces terms and Figures 6.c depicts $H_b, H_g$ and their velocities for an example droplet with $\lambda_{max}$. Figures 6.d-e compare the experimental final self-ejection velocity ($v_{ej,exp}$) of the centre of mass and transient times with the theoretical ranges ($v_{ej,th}$) for various equivalent radii, both for the viscous and not viscous cases. In Figures 6.f, the theoretical efficiency $\eta_{th}$ (Eq. 19) is reported:

$$\eta_{th} = \frac{\rho V_{ej} v_{ej,th}^2}{2\,\Delta E_{surf}} \quad (19)$$

where $\Delta E_{surf} = E_{surf,initial} - E_{surf,final}$ (Eq. 20) is the surface energy difference between the initial and final state of self-ejection from a conical pore, representing the maximum kinetic energy that the droplet would gain if there were no viscous and capillary dissipations.

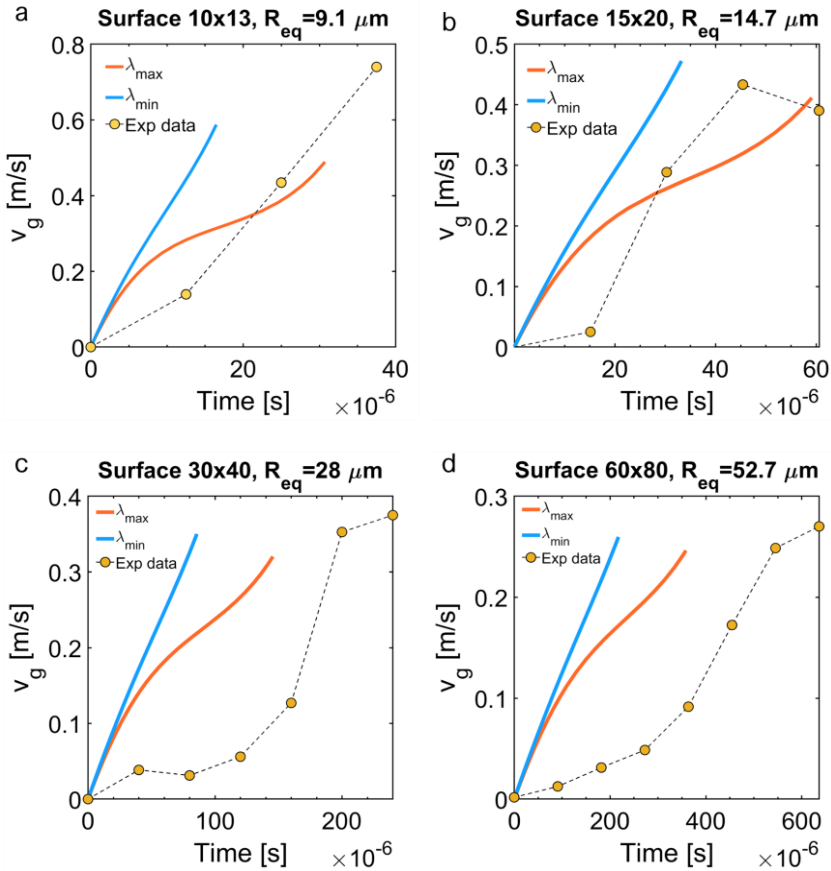

*Figure 7. Comparison between experimental (Videos S3-6) and modelled (viscous case) self-ejection transients of A) Surface 10x13, captured at 80 000 fps, B) Surface 15x20, captured at 66 000 fps, C) Surface 30x40, captured at 25 000 fps and D) Surface 60x80, captured at 11 000 fps.*

The model predicts $v_{ej}$ very well both qualitatively and quantitatively. The underestimation of the transient time, larger for larger droplets, will be discussed in the next section. The percentage of energy dissipated by capillary forces (~45÷60 %) is greater than viscous ones (~15÷25 %). In the viscous case $\eta_{th}$ is in the range of 20÷55 %, much higher than that of coalescence-induced jumping (<6%). Figure 7 compares the experimental and simulated transients. Apart from an initial delay, the experimental and theoretical velocity curves share a similar slope, thus a similar driving force. As explained in more detail in the next paragraph, the delay may be due to the different way in which the striction is overcome in the solid microcones case: not during the pre-ejection-growth but during the self-ejection itself.

### Self-ejection from microcones

The modelling of the growth phases, self-propulsion and self-ejection considering the conical pore resulted in a great simplification of the solid and especially of the droplet geometry, allowed an analytical treatment and offers a first mechanistic explanation of all the steps. The only parameters used are experimental ($\beta, \vartheta_a$ and $\vartheta_r$ of the surface) and thus the correspondence between theoretical and experimental $v_{ej}$ and self-propulsion distance indicate that the forces described approximate well the real ones. In this paragraph, we describe the drop between the four truncated cones at the beginning of the ejection transient based on video analysis, highlight the differences compared to the pore model and, by approximating the droplet with a spheroid, we estimate $v_{ej}$ with an energetic model.

Figure 8.A specifies the angles from which Videos S1-S6 (view perpendicular to the cleavage line of the samples), Video S7 (view from above) and Videos S8-9 (view at 45° to the cleavage line) were captured. Video S1-2 show that after cycles of growth and stopped self-propulsion the droplet reaches the edges of the truncated cones, $\vartheta_t$ slightly increases during the subsequent growth by condensation (without reaching 90° respect to the horizontal, which corresponds to $\vartheta_t = \pi - \beta$) and the contact radius seems to slightly enlarge laterally, on the heads of the truncated cones. In fact, Videos S8-9 show that, an instant before the beginning of self-ejection, the upper contact line is beyond the head edge without having reached $\vartheta_a$ with respect to the heads-plane but an average angle of ~74° which corresponds to $\vartheta_t = \vartheta_a$ with respect to the walls. This detail hints that in complex geometries, the drop system can assume shapes that minimise interfacial energy rather than locally respecting the dynamic angles which, indeed, are measured under very particular conditions of symmetry. This lateral growth that incorporates the microcones implies a decrease in internal pressure (the contact radius of the *top* meniscus enlarges) and $\vartheta_b$, to follow, decreases to $\vartheta_r$. Then the droplet accelerates and $\vartheta_t$ increases with fixed contact radius, as assumed with the conical pore model.

In our opinion, the differences in transient time (Figures 6.e and 7) depend on when the striction is overcome: during and before the acceleration in the real and in the modelled case, respectively. In other words, during the transient, in the modelled case the opposing force $F_{t,\sigma}^{\uparrow}$ is a decreasing function while in the real case it has a maximum when $\vartheta_t = \pi - \beta$. However, once the striction is overcome, the similar acceleration indicates a well-estimated force. This similarity can be explained as: i) although the droplet contacts the outside of four solid cones instead of the inside of a cone, the ratio between the negative ($F_{t,\sigma}^{\uparrow}$) and positive terms ($F_{b,\sigma}^{\uparrow}, F_p^{\uparrow}$) per unit length of the contact line should be roughly the same and thus also the opposing $F_{capillary}$. ii) the experimental $\lambda_{ej} \sim 1.25$ (calculated with the procedure explained in Supplementary information S1) is in the modelled range and thus also the stored surface energy is well-estimated.

Videos from various angles show a droplet resembling a spheroid intersecting the four cones. For simplicity's sake, let us consider the drop before self-ejection as a spheroid generated by the rotation of an ellipse of semi-axes $a$ and $c$ and with its centre at coordinates $(x, \hat{z}) = (\sqrt{2}p/2, h)$, as illustrated in Figure 8.b-c. Given the low eccentricity of the ellipse, $e = \sqrt{1 - a^2/c^2}$, we approximate the curvatures of the *top* and *bottom* menisci as being those of spherical caps, $R_t = -(\sqrt{2}p - d_t)/[2\cos(\vartheta_a + \beta)]$ and $R_b = -(\sqrt{2}p - d_t - 2 \cdot \Delta H \tan \beta)/[2 \cos(\vartheta_r - \beta)]$, respectively, where the contact angles are the dynamic ones because we are considering the instant before the transient. Under the assumption of uniform internal pressure, $R_t = R_b$ and thus:

$$\Delta H = \frac{\sqrt{2}p - d_t}{2\tan\beta}\left[1 - \frac{\cos(\vartheta_r - \beta)}{\cos(\vartheta_a + \beta)}\right] \tag{20}$$

The spherical caps heights are $h_t = R_t[1 - \cos(\vartheta_a + \beta - \pi/2)]$ and $h_b = R_b[1 - \cos(\vartheta_r - \beta - \pi/2)]$, thus $2c = \Delta H + h_t + h_b$. By imposing the passage of the generic ellipse, $(x - \sqrt{2}p/2)^2/a^2 + (\hat{z} - h)^2/c^2 = 1$, through points $A(d_t/2, d_t/(2\tan\beta))$ and $B(d_t/2 + \Delta H \tan\beta, d_t/(2\tan\beta) + \Delta H)$, we find $h$ and $a$. The surface area and the volume of a spheroid are $A_{spheroid} = 2\pi a^2(1 + \frac{c}{a \cdot e}e)$ and $V_{spheroid} = 4\pi a^2 c/3$. The radius of a sphere of the same volume is $R_{eq,th} = [3V_{spheroid}/(4\pi)]^{1/3}$. If the difference in surface energy between the initial and final states (Eq. 21) was fully converted into kinetic energy, the droplet would have the self-ejection velocity expressed in Eq. 22. With the experimental ejection velocity ($v_{ej,exp}$) and volume $V_{exp}$, we express the efficiency of self-ejection from truncated cones as $\eta$ (Eq. 23). We report the experimental data and model results for the four surfaces (Table 2), compare $v_{ej,exp}$ with $v_{ej,th}$ (Figure 8.d*)* and plot $\eta$ (Figure 8.e) for the various self-ejection radii. An indication of the goodness of the model in estimating the droplet geometry is the average relative error of $R_{eq,th}$ compared to $R_{eq,exp}$ of 0.1 %. Figure 8.d shows that the model follows the experimental trend well and, as we expect, it overestimates the final ejection velocity. In fact, the total conversion of $\Delta E_{surf}$ to kinetic energy neglects dissipations due to adhesion, capillary and viscous dissipations which, as seen in the above section, introduce a loss $(1 - \eta_{th})$ of 50 to 80 %. In Figure 8.e $\eta$ roughly confirms experimentally the efficiency values (around 50 %) predicted by the modelled transient in the conical pore. The case of *Surface 10x30* is an exception and shows a higher efficiency (69 ± 29 %).

*Table 2. Experimental data and modelling results for the droplets on the four surfaces.*

| Surface | $a_{exp}$ [µm] | $a$ [µm] | $c_{exp}$ [µm] | $c$ [µm] | $R_{eq,exp}$ [µm] | $R_{eq,th}$ [µm] | $v_{ej,exp}$ [m/s] | $v_{ej,th}$ [m/s] | η |
|---|---|---|---|---|---|---|---|---|---|
| 10x13 | 8 | 8.04 | 10.6 ± 0.3 | 10.57 | 9.1 ± 0.4 | 8.81 | 0.65 ± 0.13 | 0.81 | 0.69 ± 0.29 |
| 15x20 | 13 | 12.78 | 15.9 ± 0.1 | 16.83 | 14.0 ± 0.5 | 14.01 | 0.44 ± 0.06 | 0.65 | 0.46 ± 0.13 |
| 30x40 | 27.5 | 26.47 | 33.0 ± 0.6 | 34.85 | 28.8 ± 1.7 | 29.01 | 0.31 ± 0.06 | 0.45 | 0.46 ± 0.19 |
| 60x80 | 48.5 | 49.03 | 59.3 ± 0.9 | 64.57 | 52.7 ± 1.2 | 53.74 | 0.25 ± 0.02 | 0.33 | 0.52 ± 0.09 |

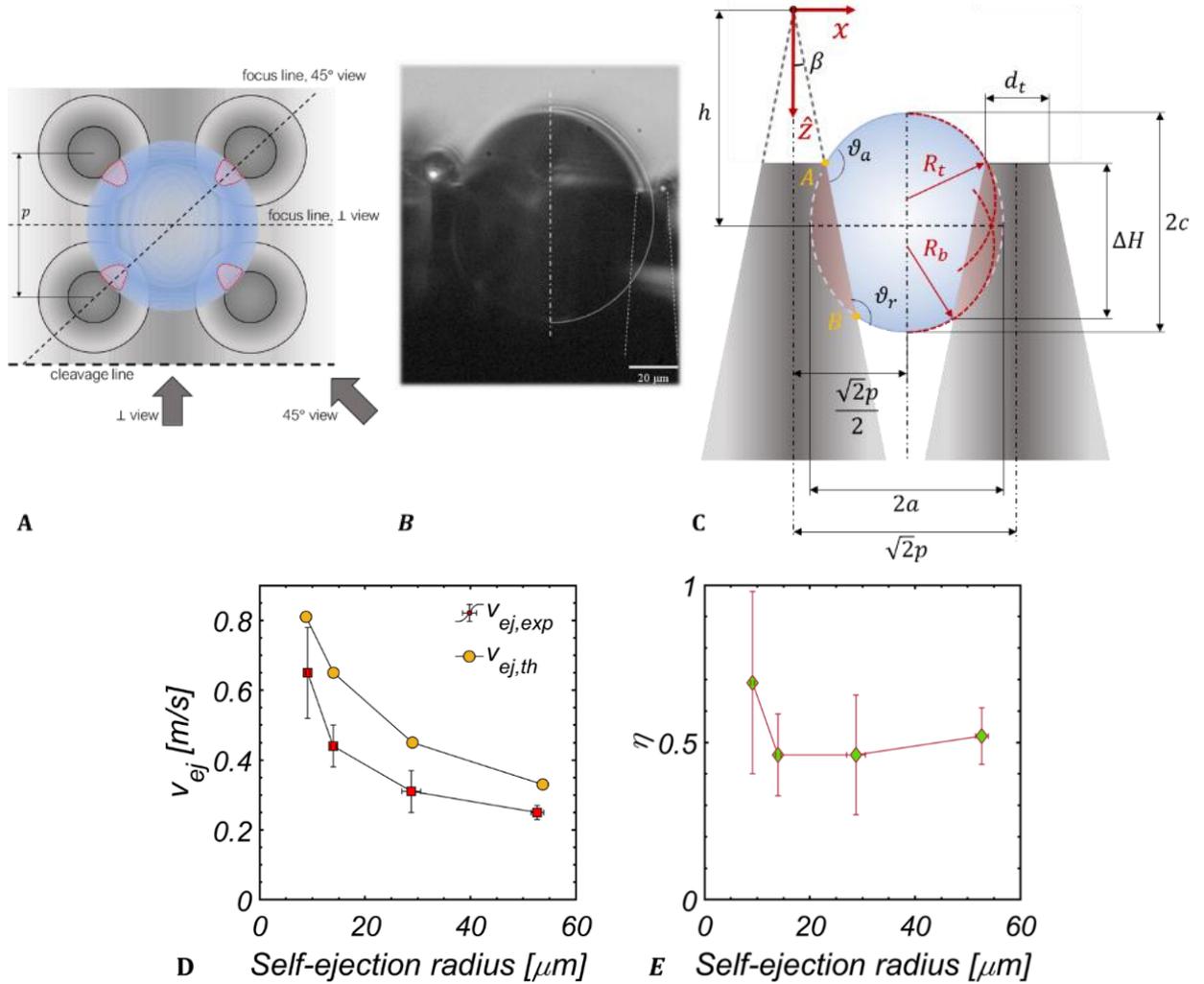

*Figure 8. A) Diagram of the surface with the 3 views used in the experiments: from above (view of Figure 8.A itself), perpendicular to the silicon cleavage line and at 45° to that line. The planes of focus pass through the centre of the droplet. B) Frame of Video S8, 45° view, highlighting the elliptical section of the drop. C) Dimensioned diagram of the drop from the 45° view used for the spheroid model developed in Section 4. D) Comparison of $v_{ej,exp}$ and $v_{ej,th}$ for various self-ejection radii ($R_{eq}$). E) Self-ejection efficiency with propagated error bars.*

$$\Delta E_{surf,cones} = \sigma_{lv}(A_{spheroid} - 4\pi R_{eq,th}^2) \tag{21}$$

$$v_{ej,th} = \sqrt{\frac{2\Delta E_{surf,cones}}{\rho V_{spheroid}}} \tag{22}$$

$$\eta = \frac{\rho V_{exp} v_{ej,exp}^2}{2 \Delta E_{surf}} \tag{23}$$

## Discussion

To model the mechanism of growth and eventual self-motions of individual condensation droplets in divergent structures with uniform wettability, we have chosen the conical pore as a case study. This choice allows general insights and is easier to model analytically than the wedges reported elsewhere or the solid cones of our experimental study, due to its axial symmetry and simple geometry of the contact areas. We described the external forces acting on the system "*droplet suspended in a conical pore with hysteresis*" and verified that the resultant is null at uniform internal pressure, that is, if Eq. 3 stands. In the case of a droplet with a certain initial shape ratio, we considered that it evolves through iso-pressure configurations as it is free to accommodate the increasing volume by varying the contact angles and contact line heights. Therefore, as long as both menisci are not in a dynamic configuration, the drop can be described by Eq. 3 alone and we identified two growth phases. Once the dynamic configuration is reached, the droplet is free to move and release interfacial energy; the propulsive surface force and the opposing ones (capillary and viscous) determine its motion. By solving its dynamics, we predicted that it rapidly self-propels, then stops at a distance dependent on the surface parameters, it adjusts the contact angles to cancel internal pressure gradients and continues to slowly grow. After a certain number of growth and self-propulsion cycles, it reaches the edge of the opening. It grows to one of the possible dynamic configurations and self-ejects. In this study, we investigated the effect of the droplet size on self-ejection velocity by fixing certain taper and contact angles. To test the theory but also to avoid air depressurization problems that the pore may introduce during the fast movements, we designed uniformly hydrophobic, nanostructured truncated microcones arranged in a square pattern. A droplet in between four of them is analogous to the modelled one in terms of the forces at play. After the development of a t-RIE recipe for Si etching and the investigation on the effect of the lithographic mask geometry on the structures, we fabricated four arrays of truncated microcones with the same tapering but different size. Then, the nanostructuration of deposited Al and silanization rendered the microstructures highly and uniformly hydrophobic (free of strong pinning sites, at least not appreciable in the experiments). The observations of the condensation droplets with high spatial and temporal resolution confirmed the growth-propulsion cycles and reveal for the first time the self-ejection of single droplets on this type of surfaces and its dynamics. The self-ejection radius is precisely controlled by the structure geometry differently from coalescence-jumping where it is a random variable. The models are in good agreement with the observed self-ejection transients and final ejection velocity, decreasing with droplet radius. A slight underestimation of the transient time is argued on the basis of the conical pore and solid cones differences and of the analysed videos. In addition, we also propose an energetic model of the droplet between the four truncated cones for a rapid estimate of the final self-ejection velocity. The self-ejection efficiency estimated with our models is around 50%.

In future theoretical and experimental studies we aim to investigate self-ejection as dynamic contact angles and cone geometry vary. By preliminary evaluations with our model, we expect lower limit values of the receding angle and upper limit values of the tapering beyond which the self-ejection does not occur. The present research describes with a novel approach, and proves experimentally, the self-ejection in its essentiality, without the need for pinning sites and therefore also with the benefit of a facilitated fabrication. However, due to similar conditions, it will be possible to adapt our model to the case of self-propulsion and/or ejection caused by an abrupt change of a contact angle (for example, the detachment from an inserted pinning site/bottom of structures) reported elsewhere.

Moreover, we deduce that ideal cone arrays (with a sharp tip) would nullify the percentage of droplets that nucleate on the heads and can only leave the surface by coalescence jumping. In this way, all the droplets would leave the surface by self-ejecting at a precise, designed size without the variability typical of coalescence jumping. To obtain this result on a surface structured with rectangular microgrooves would mean to reduce the wall thickness towards zero with the consequent instability and poor mechanical resistance. Microcones with a small heads area fraction, instead, would maintain mechanical resistance, higher for larger $\beta$. Future studies will explore the optimal $\beta$ that allow self-ejection and provide the highest mechanical resistance.

In conclusion, the uniformly hydrophobic divergent structures reported enable self-ejection of droplets with precise size and higher efficiency compared to coalescence-jumping, permit an easier fabrication compared to biphilic structures and add a piece to the knowledge on the self-motion mechanisms. This new class of surfaces will bring further enhancements in the applications where coalescence jumping has already introduced improvements.

## Materials and Methods

### Microstructures fabrication

We employed 6-inch silicon wafers (100) as the substrate to fabricate microcones by photolithography and tapered reactive ion etching (t-RIE) (Figure 9.a-b). After a standard RCA cleaning, the hard mask was made by growing 200 nm of thermal silicon oxide (Centrotherm E1200HT furnace) followed by the deposition of 200 nm of Aluminium by magnetron sputtering (Eclipse MRC). Then we deposited 1.2 μm of positive photoresist by spin coating (Track SVG). We employed two photolithographic masks (Photronics):

1) Mask1 consists of areas of 1 cm x1 cm with arrays of circles arranged either in a square or hexagonal pattern, each with a different circle diameter ($D_{mask}$) and pitch ($p$) (green areas in Table S1). It was designed to investigate the effect of the non-masked area fraction, of $D_{mask}$, of the pattern type and of the etching time on the t-RIE in terms of microstructures tapering, etch rate and surface characteristics such as the roughness of the walls and the presence of grass on the bottom. The patterns are transferred on the photopolymer-coated wafers by one-step uv-light exposure (mask aligner MA150CC).

2) After the studies with Mask1, we designed Mask2 with circles arranged in a square pattern with optimal $D_{mask}$ and $p$ (Table 1). The square pattern allows condensation droplets to be observed against high-intensity backlighting, a crucial aspect for acquisitions at high frame rates (see paragraph

Condensation experiments). Mask2 is for use with a stepper photolithographic machine (Nikon 2205i11D) and its pattern was reproduced on large areas, as depicted in Figure S7 , by step-and-repeat exposure. Working over large areas facilitates manipulation in the subsequent phases and in particular the cleavage with a breaking line passing over the pattern and parallel to the rows of cones.

After developing (Track SVG), hard bake of the photo-polymer was carried out. The pattern-transfer onto the hard mask was performed by dry etching of Aluminium (KFT Metal PlasmaPro100 Cobra300) and silicon oxide (Tegal 903e). Then the t-RIE step was performed (Alcatel dry etcher). The scallops typical of the Deep Reactive Ion Etching (DRIE) [60,61] may be strong pinning sites for droplets and frustrate the self-ejection. Therefore, we opted for continuous etching[62,63] using $SF_6$-$C_4F_8$ plasma[64], a process without scallops and in which the ratio of gas flows, chamber pressure, bias and source power and temperature influence the tapering and uniformity of the sidewalls. We developed a t-RIE recipe with the purpose of creating truncated microcones with tapering angle ($\beta$, see Figure S3.a) in the range 5÷10° and as similar as possible among the four arrays different in size to be tested in condensation conditions. Also, the etch rate has to be preferably higher than 500 nm/min and the lateral and bottom walls as smooth as possible. The recipe parameters are: source power 2800 W, bias power 20 W, gas fluxes ratio $SF_6/C_4F_8$=0.65, total gas flux 500 sccm, chamber pressure 0.04 mbar and wafer temperature 20 °C. The evidences of the campaign of experiments carried out with Mask1 are: $\beta$ is the result of vertical etching and horizontal etching under the mask and is constant from the bottom base up to about three fourth of the microstructure, then it goes to zero on top; $\beta$ and the etch rate have a peaked trend with the etching time, for each particular mask geometry; $\beta$ has either a decreasing or a peaked trend with the not-masked area fraction $\varphi$; for $\varphi$>0.7, irregularities of the lateral and bottom walls of the microstructures (microribs) appear; the mean etch rate is 680 ± 80 nm/min. In S5.1 (Supplementary information) we report the measurement method of $\beta$ and images and plots in support of the mentioned trends.

Then we fabricated the four patterns (Mask2) to test in condensation conditions, designed with $D_{mask}$x $p$ optimal to minimize walls irregularities and with similar $\beta$ (Figure 1.c-f and Table 1). This $\beta$ is a medium one relative to the upper part of the microstructures involved in self-ejection (see S5.1, Supplementary information where it is called $\bar{\beta}$). It is on average the same for the four surfaces. We removed the etching passivation layer by immersion in isopropanol with ultrasonic pulses and then the hard mask by dipping the wafers in an Al etch solution and then in a Piranha solution. We cleaned them in a deionized water rinse until the bath reached 16 MΩ-cm. The truncated microcones have a variating undercut at the apex. We removed it with isotropic etching (Tegal 900) which lowered the pillars by about 2 μm and made the top part straight (see S5.1, Supplementary information).

### Nanostructuring, silanization and wettability

As a second hierarchical level we have selected the aluminium to be nanostructured (NanoAl) with HWT as it is compatible with clean room processes, the structuring is simple, cheap, scalable and could also be used for microstructures produced directly on aluminium through other industrial processes. The hot water treatment of many metals and their alloys leads to the formation of nanostructures. A thin superficial layer of metal oxide forms in hot water, the oxide cations are released in solution, migrate and deposit forming nanostructures with peculiar shapes for each metal[65,66]. In the case of Al, thin nano-blades of hydrated aluminium oxide (pseudo-boehmite) are formed which, once made hydrophobic, have shown superhydrophobic and anti-freezing properties[67–73]. Both water temperature and treatment time have an effect on contact angles[70,74,75]. We deposited 150 nm of pure Al on the wafers by e-beam evaporation (ULVAC HIGH VACUUM COATER EBX-16C) (Figure 9.c). The wafers were cleaved as shown in Figure S7 in order to have samples with a row of cones on the sharp edge. HWT was performed by immersion in deionized water (18 MΩ-cm) at 90 °C for 7 min (Figure 9.D), immediately followed by immersion in deionized water at room temperature to block the structuring and then dried with a $N_2$ flow. The SEM image of the nanostructures in Figure 1.b was done with a Tescan SEM tool. We cleaned the surfaces by dipping in acetone, isopropanol and deionized water, dried with a $N_2$ flow and activated the surfaces with oxygen plasma (to increase the amount of silanols and maximize the uniformity and density of the self-assembled monolayer[76]). Chemical vapour deposition (CVD)[77–80] of 1H,1H,2H,2H-Perfluorodecyltriethoxysilane (Sigma-Aldrich) was performed by placing the samples and 200 μl of fluorosilane in a sealed (class IP-67) aluminium box (internal volume of 3.7 litres) heated at 150 °C for 3 h followed by an annealing for 1.5 h with the box opened (for covalently unbound silane removal).

To characterize the dynamic contact angles of water on the walls of the microcones we measured them on the non-microstructured areas of the samples, only covered by NanoAl (Figure S). We characterized them with two procedures (Figure S9):

1) Macro-droplets: we employed a syringe pump (Pump 11 Elite, Harvard apparatus) to inject and aspirate a droplet of deionized water at a volume rate of 3 μl/min (to avoid dynamic effects[81]) through a syringe with diameter 230 μm (gauge 32), positioned close and perpendicular to the surface. We captured the two steps with a digital microscope (Dinolite AM7915MZTL) and measured the dynamic contact angles with DropSnake[82], an ImageJ plug-in. The values were acquired when the droplet has a diameter at least five times the syringe diameter: $\vartheta_a$=166 ±1° and $\vartheta_r$=123 ±7°.

2) Micro-droplets: we captured the micro droplets (tens of μm) during condensation and evaporation with the experimental setup described in the next paragraph (Phantom camera + microscopy objective) and analysed with DropSnake. $\vartheta_a$=157 ±1° and $\vartheta_r$=145 ±6. We used the micro-droplets contact angles for the analytical models as they are characteristic of the droplets affected by growth, self-propulsion and self-ejection and account for the capillarity alone. The macro-droplets angles are instead influenced by the gravitational force.

The equilibrium contact angle of NanoAl can be estimated using the experimental dynamic angles[83] with $\vartheta_{eq,NanoAl} = \cos^{-1}[(\cos\vartheta_a + \cos\vartheta_r)/2] \approx 150°$. A theoretical estimate can be obtained by applying the Cassie-Baxter equation[84], $\cos\vartheta_{CB} = f(\cos\vartheta_{eq,flat} + 1) - 1$. With the micro-droplets procedure on a flat Si surface, covered by e-beam evaporated Al and silanized, we measured $\vartheta_{a,flat}$=111.4 ± 0.2° and $\vartheta_{r,flat}$ =89.8 ± 1.5°; thus, $\vartheta_{eq,flat} \approx$100°. With ImageJ's *Particle Analysis* plug-in we roughly quantified the area fraction of the solid-liquid interface ($f$ ~0.17) (Figure S10), thus $\vartheta_{CB} \approx$149°. The good agreement with $\vartheta_{eq,NanoAl}$ suggests that droplets are in a fakir state on NanoAl.

Condensation experiments

We performed the condensation experiments in a custom-made setup (see Figure 1.k for the setup scheme) according to the following procedure: we introduced the desired humid air in a chamber (800 cm³) by mixing a dry and a wet air flux. The wet air flux is obtained by passing dry air in a bubbler filled with deionized water (18 MΩ-cm). Each flux is set with a flow meter (FR2000, Key instruments) and the total flux is 800 sccm. The relative humidity ($RH$) and temperature of the mixed air ($T_a$) are measured with an Arduino BME280 sensor (accuracies ±3% and ±1 °C) placed at the chamber inlet. The cold plate inside the chamber is cooled by a thermostatic bath and two Peltier stages to $T_p$= 1°C (measured with a thin film PT100 thermocouple, RS pro, class B accuracy) for the duration of the experiments. The sample and the PT100 are in thermal contact with the cold plate through a thermal pad (T-flex 600 Series Thermal Gap Filler, Laird Technologies, thickness of 1 mm, thermal conductivity of 3 W/mK). Given the low thermal inertia of the silicon samples (thick 600 μm) and the PT100, we assume surface temperature $T_{surf} = T_p$. The water vapour pressure ($P_{vap}$) of the fluxed air is 11.8 hPa thus the saturation ratio on the sample surface is $s = P_{vap}/P_{vap,sat}(T_{surf})$=1.8, where $P_{vap,sat}(T_{surf})$ is the saturation vapour pressure at $T_{surf}$. We placed the sample on the plate, introduced dry air and cooled to $T_{surf}$=1 °C. Then, humid air is introduced and condensation starts. The chamber is equipped with an upper and a lateral quartz window. To observe condensation (and evaporation, for the micro-droplet contact angle measurements) we employed a high frame rate camera (Phantom V640, Vision Research) coupled to a microscopy objective (50X Mitutoyo Plan Apo infinity corrected, long working distance=13 mm, resolving power=500 nm, depth of focus=900 nm) through a tube lens (InfiniTube Ultima). We illuminated the surfaces with a LED light (MULTILED QT, GSVITEC) placed outside the environmental chamber and on the back of the samples with respect to the video camera. The growth and self-propulsion videos were captured at 24÷2000 fps while self-ejection transients at 9000 ÷ 80000 fps. The transients frames were analysed in ImageJ to measure the position ($z_g$) evolution of the centre of mass (*g*) considered as the centre of a fitted ellipse (Figure 1.l). The velocity at the instant $i$, $v_g(i)$, is calculated as $[z_g(i) - z_g(i-1)]/\Delta t$, with $\Delta t$ the time length of a frame. A 3-axis micro positioning stage was used to move the camera and to focus.

## Acknowledgements
We kindly thanks Alberto Bellin, Luigi Fraccarollo and Fabio Sartori (University of Trento, Italy) who lent us the Phantom camera, Elia Scattolo (FBK, Trento, Italy) for the SEM images of the nanostructures, the Micro Nano Facility technical staff (FBK, Trento, Italy) for support in the fabrication and Stefano Siboni and Claudio Della Volpe (University of Trento, Italy) for fruitful discussions. The study has been supported by the European Commission under the FET Open "Boheme" grant no. 863179.

## Author contributions
A.B. and N.M.P. supervised the research. N.G.D. and A.B designed the study, designed, fabricated and tested the surfaces and interpreted the data. N.G.D. developed the modelling and wrote the manuscript. A.B. and N.M.P. discussed the modelling and edited the manuscript.

# Supplementary information

## S1. Equilibrium of a droplet suspended in a conical pore with uniform internal pressure

We here proof that, if the internal pressure is uniform (Eq.3, here rewritten as Eq. S1.1), then the external forces have a null resultant.

$$H_t = H_b \frac{\cos(\vartheta_t + \beta)}{\cos(\vartheta_b - \beta)} \qquad \text{S 1.1}$$

Considering Figure 2.a, by substituting the external forces (Eqs. 3-6) in $F^\uparrow_{capillary} = F^\uparrow_{t,\sigma} + F^\uparrow_{b,\sigma} + F^\uparrow_p$, after some manipulations, we obtain:

$$F^\uparrow_{capillary} = 2\pi\sigma_{lv} H_b \tan(\beta) \cdot \left[ \frac{\cos^2(\vartheta_t + \beta)}{\cos(\vartheta_b - \beta)} - \cos(\vartheta_b - \beta) - \frac{\cos^2(\vartheta_t + \beta) - \cos^2(\vartheta_b - \beta)}{\cos(\vartheta_b - \beta)} \right] = 0 \qquad \text{S 1.2}$$

which is identically zero for each combination of parameters.

## S2. Calculations of the shape ratio $\lambda_{max}$

The analysis of the shape ratio before self-propulsion was made on six self-propulsion events captured from the orthogonal view. As explained in detail in S5, the tapering of our solid microcones is constant for about three fourth from the bottom and then they gradually becomes straight near the top. To calculate the theoretical $\lambda_{max}$ with Eq.3, since the droplets captured are about at a middle height, we employ the constant tapering of the two thirds of the cone. Given the slightly different $\beta$ of the four type of cones, we used an average $\beta$=7.8±1.3 and, accounting also for the standard deviation of the dynamic contact angles, we obtained $\lambda_{max}$=1.31±0.13, where the standard deviation is the propagated error.

The measurement of the experimental $\lambda_{max,exp}$ is done by assuming that the droplet between the four cones is similar to the one in the pore, as illustrated in Figure S1. However, it is not direct: we captured the droplets from the orthogonal view which allows to well-analyse the droplet shape with the light on the back but not the real points of contact. Those are only visible from the 45° view which is, however, problematic for illumination because light is blocked by the cones behind. Moreover, compared to the pore model, we do not directly measure $H_b$ and $H_t$ but $h_b$, $h_t$ and $x_\perp$. Eqs. S2.1-2 show how to calculate $H_b$ and $H_t$, in reference of Figure S1. $\lambda_{max,exp}$=1.19±0.02.

$$H_b = \frac{h_b + \frac{2x_\perp + p(\sqrt{2} - 1)}{2\tan\beta}}{1 + \frac{\tan\beta\,[1 - \sin(\vartheta_r - \beta)]}{\cos(\vartheta_r - \beta)}} \qquad \text{S 2.1}$$

$$H_t = \frac{h_t + \frac{2x_\perp + p(\sqrt{2} - 1)}{2\tan\beta}}{1 - \frac{\tan\beta\,[1 - \sin(\vartheta_a + \beta)]}{\cos(\vartheta_a + \beta)}} \qquad \text{S 2.2}$$

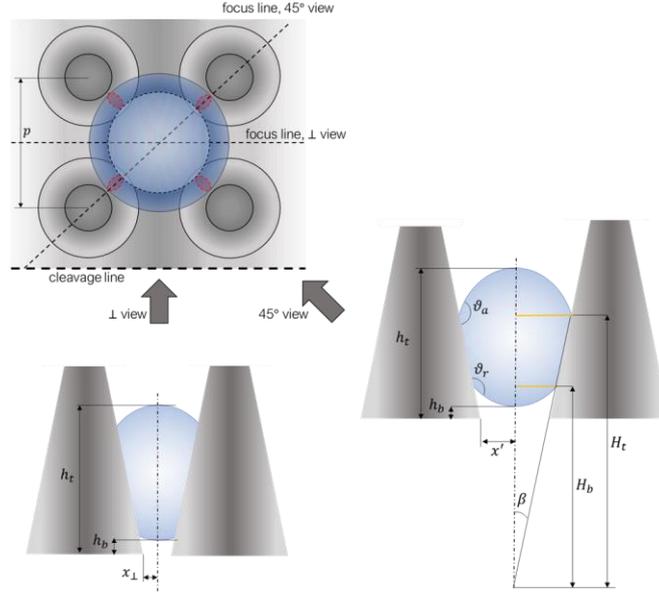

*Figure S1. Scheme of the views. Videos are captured from the orthogonal view while the 45° view is the one to consider to apply the conical pore model.*

## S3. Expressions and relations between variables in self-propulsion and self-ejection

### S 3.1 Volumes and surface energy

The volume of the top and bottom spherical caps and of the truncated cone are:

$$V_{cap,t} = \frac{\pi}{3}\left(\frac{-H_t \tan\beta}{\cos(\vartheta_t + \beta)}\right)^3 (2 + \sin(\vartheta_t + \beta))(1 - \sin(\vartheta_t + \beta))^2 \tag{S 3.1}$$

$$V_{cap,b} = \frac{\pi}{3}\left(\frac{-H_b \tan\beta}{\cos(\vartheta_b - \beta)}\right)^3 (2 + \sin(\vartheta_b - \beta))(1 - \sin(\vartheta_b - \beta))^2 \tag{S 3.2}$$

$$V_{tc} = \frac{\pi}{3} \tan^2\beta \left(H_t^{\ 3} - H_b^{\ 3}\right) \tag{S 3.3}$$

The surface energy of the droplet is:

$$E_{surf} = \sigma_{lv}(A_{cap,t} + A_{cap,b} - A_{truncated\ cone} \cos\vartheta_{eq}) =$$
$$= \sigma_{lv}\{2\pi R_t^2[1 - \sin(\vartheta_t + \beta)] + 2\pi R_b^2[1 - \sin(\vartheta_r - \beta)] - \pi(H_t^{\ 2} - H_b^{\ 2})\cos\vartheta_{app} \tan\beta / \cos\beta\} \tag{S 3.4}$$

Depending on which motion is simulated, $\vartheta_t$ and $H_t$ are fixed or variable as described in the manuscript.

### S 3.2 Self-propulsion

The general expression of the volume $V$ of a droplet composed by two spherical caps and a truncated cone is Eq.9. In the case of self-propulsion, the droplet reaches $\lambda_{max}$ so by substituting $\vartheta_b = \vartheta_r$, $\vartheta_t = \vartheta_a$, $H_b = H_{b,0}$ (arbitrary value) and $H_t = \lambda_{max} H_{b,0}$ in Eq. 9 we calculate $V^*$, the volume during self-propulsion of a droplet with an initial $H_{b,0}$. The volume is reasonably constant because during transients (tens to hundreds of μs) the condensation volume is negligible and we assume water to be uncompressible. Therefore, rearranging $V = V^*$ (Eq. 10), we obtain $H_t$ as a function of $H_b$:

$$H_t = \sqrt[3]{\frac{V^* + \frac{\pi}{3}H_b^{\ 3}\left\{\tan^2\beta + \left(\frac{\tan\beta}{\cos(\vartheta_r - \beta)}\right)^3 [2 + \sin(\vartheta_r - \beta)][1 - \sin(\vartheta_r - \beta)]^2\right\}}{\frac{\pi}{3}\left\{\tan^2\beta - \left(\frac{\tan\beta}{\cos(\vartheta_a + \beta)}\right)^3 [2 + \sin(\vartheta_a + \beta)][1 - \sin(\vartheta_a + \beta)]^2\right\}}} \tag{S 3.5}$$

In general, the centre of mass of a spherical cap with respect to the centre of the sphere and the one of a truncated cone with respect to the larger base are expressed in Eq. S 3.6 and Eq. S 3.7, respectively.

$$z_{cap} = \frac{3}{4}\frac{(2R - h_{cap})^2}{3R - h_{cap}} \tag{S 3.6}$$

$$z_{tr} = \frac{h_{tr}}{4}\frac{(r_t^2 + 2r_t r_b + 3r_b^2)}{r_t^2 + r_t r_b + r_b^2} \tag{S 3.7}$$

$h_{cap}$ and $h_{tr}$ are the height of the cap and of the truncated cone, respectively. The position of the centres of mass with respect to the apex of the cone are:

$$Z_{cap,t} = H_t - (R_t - h_{cap,top}) + z_{cap,top} \tag{S 3.8}$$

$$Z_{cap,b} = H_b + (R_b - h_{cap,bottom}) - z_{cap,bottom} \tag{S 3.9}$$

$$Z_{tc} = H_t - z_{tr} \tag{S 3.10}$$

By substituting Eq. 10, Eqs. S 3.1-3, Eqs. S 3.5-10, $\vartheta_b = \vartheta_r$ and $\vartheta_t = \vartheta_a$ in Eq. S 3.11 one obtains the position of the centre of mass of the droplet with respect to the apex of the cone, $H_g$, as a function of $H_b$ alone.

$$H_g = \frac{Z_{cap,t}V_{cap,t} + Z_{cap,b}V_{cap,b} + Z_{tc}V_{tc}}{V^*} \tag{S 3.11}$$

### S 3.3. Position, velocity and acceleration of the centre of mass

To speed up the resolution of Eq. 17, we approximated $H_g$ (Eq. S 3.11) with a sixth-order Taylor series expansion around $H_{b,0}$ in the variable $H_b(t)$, called $H_{g,T}$. The expressions of the centre of mass velocity and acceleration are obtained by derivation of $H_{g,T}$ with respect to time.

### S 3.4. Final configuration of self-propulsion

As described in the *self-propulsion Section* in the article, the droplet stops at a certain $H_{b,fin}$ to which a certain $H_{t,fin}$ corresponds, calculated with Eq. S3.5. The contact angles are still $\vartheta_b = \vartheta_r$ and $\vartheta_t = \vartheta_a$ but there is no uniform internal pressure. The contact angles rearrange to diminish the pressure difference; $\vartheta_t$ decreases and $\vartheta_b$ increases. Given the shape ratio $H_{t,fin}/H_{b,fin}$ (Figure S2) and the volume $V^*$, we search for a combination of $\vartheta_t$ and $\vartheta_b$ that verify Eq.3. We find the solution by solving numerically the system of equations: $\frac{H_{t,fin}}{H_{b,fin}} = \frac{\cos(\vartheta_t + \beta)}{\cos(\vartheta_b - \beta)}$ (Eq. 3) and Eq. 10. Real solutions mean there is an equilibrium configuration which corresponds to what we found for the surface parameters analysed. Imaginary solutions mean the contact angles change to $\vartheta_b = \vartheta_a$ and $\vartheta_t = \vartheta_b$ without finding an iso-pressure configuration; thus, $F_{capillary}^{\uparrow}$ is still negative and the droplet should accelerate in the negative direction, driven by $F_{net}^{\uparrow} - F_{visc}$. For the surface parameters of the present study we found that an equilibrium configuration can be attained.

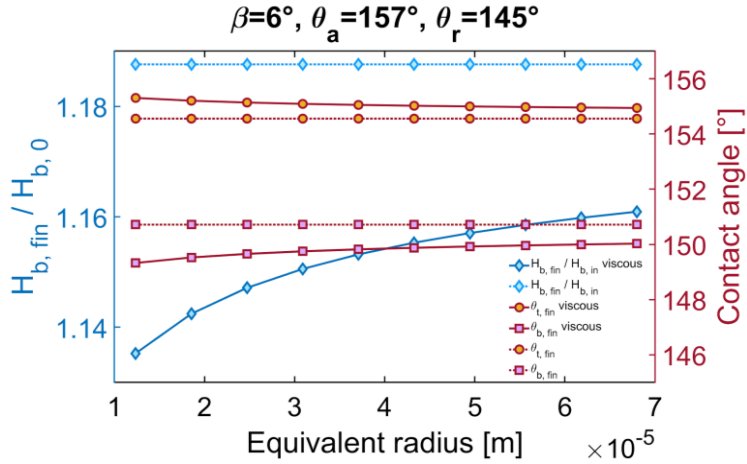

*Figure S2. $H_{b,fin}/H_{b,0}$, $\vartheta_{t,fin}$ and $\vartheta_{b,fin}$ vs $R_{eq}$ for both the viscous and non-viscous cases and for fixed surface parameters.*

## S 4. Self-ejection

In the case of self-ejection, by substituting $H_t = H_{t,max}$, $H_{b,0} = H_{t,max}/\lambda_{ej}$, $\vartheta_b = \vartheta_r$ and $\vartheta_t = \hat{\vartheta}_t$ in Eq. 9 we calculate $V_{ej}$. During the ejection, the variables that change are $\vartheta_t$ and $H_b$. Again, $V = V_{ej}$ (Eq. 10) gives a relation between $\vartheta_t$ and $H_b$. We employed the MATLAB Symbolic Toolbox to express $\vartheta_t$ as a function of $H_b$. The procedure to express $H_g$ as a function of $H_b$ and for the resolution of Eq.17 is then analogous to the one reported in the section S3.

## S 5. Fabrication

### S 5.1. Lithography and t-RIE

We tested the developed t-RIE recipe on various wafers (all the splits of Table S1, both for square and hexagonal patterns) at different etch times and observed the microcones with a Tescan SEM tool by tilting the wafer (tilt angle $\alpha$). The acquired images allow to judge the pillar status (uniformity of the tapering, over-attack with mask removal, smoothness of the walls and bottom of the structures) and measure the tapering. In some cases, the tapering is not constant over the entire height of the microcones: up to about three fourth from the bottom the tapering is constant (and it is the one shown in Figure S3.a, measured and plotted in Figure S5) while for the remaining top part the pillar gradually becomes straight or with a slight undercut. The methodology for the tapering angle calculation is shown in Figure S3.a. With the diameter of the base ($D$), of the head ($d_h$) and the projected height of the cone ($c$, the height observed with the SEM), the real tapering is expressed in Eq. S 5.1. We interestingly observed that $\beta$ varies with the not-masked area fraction (called *free area fraction* $\varphi$), indicated in Eq. S 5.2-3 for square and hexagonal patterns, respectively.

Table S1. Green areas indicate the splits of Mask1, both for square and hexagonal patterns.

|  |  | D [μm] | | | |
|---|---|---|---|---|---|
|  |  | 5 | 10 | 15 | 20 |
| p [μm] | 10 | ■ |  |  |  |
|  | 15 | ■ | ■ |  |  |
|  | 20 | ■ | ■ | ■ |  |
|  | 25 | ■ | ■ | ■ | ■ |
|  | 30 |  | ■ | ■ | ■ |
|  | 35 |  |  | ■ | ■ |
|  | 40 |  |  |  | ■ |

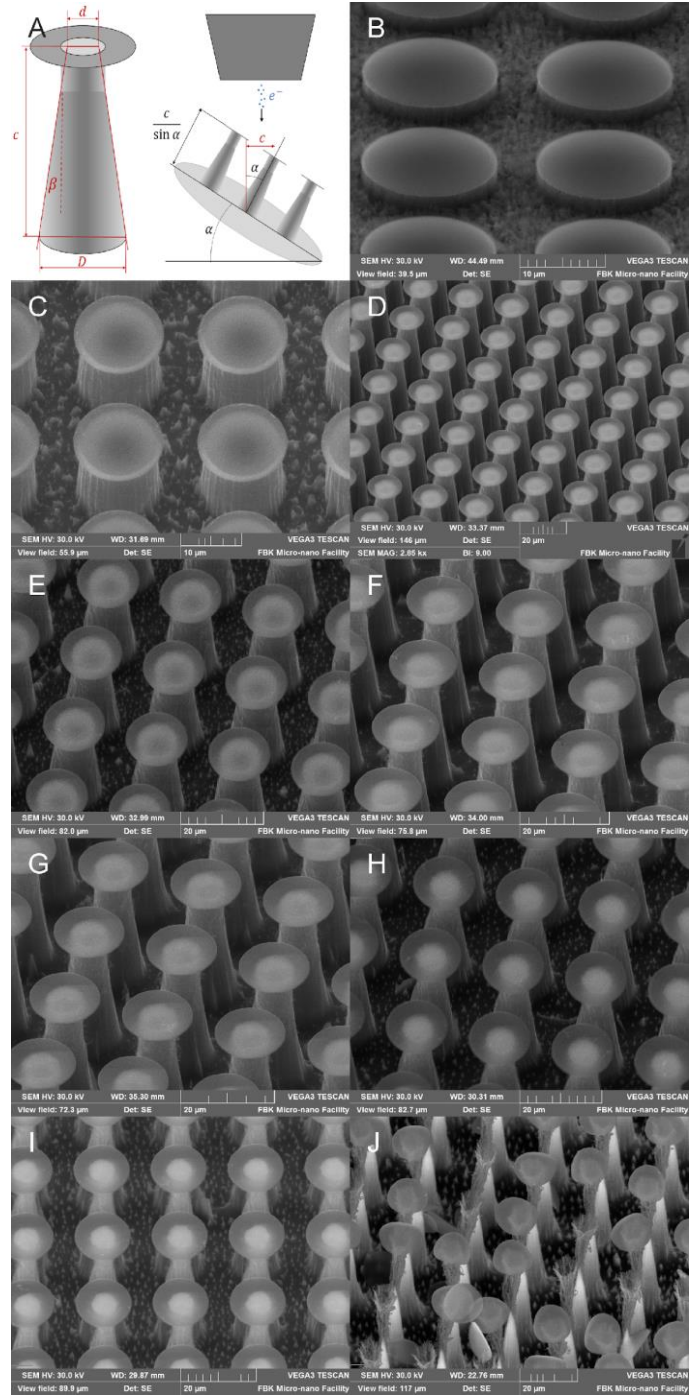

*Figure S3. A) Illustration of image analysis for the tapering calculation. B-J) SEM images of the square pattern diameter x pitch=15x20 for increasing etching times correspondent to the grey dots in Figure 4.4.*

$$\beta = \tan^{-1}\left(\frac{D - d_h}{2\frac{c}{\sin \alpha}}\right) \quad \text{(S 5.1)}$$

$$\varphi_s = 1 - \frac{\pi}{4}\left(\frac{D_{mask}}{p}\right)^2 \quad \text{(S 5.2)}$$

$$\varphi_h = 1 - \frac{\pi}{4 \sin 60°}\left(\frac{D_{mask}}{p}\right)^2 \quad \text{(S 5.3)}$$

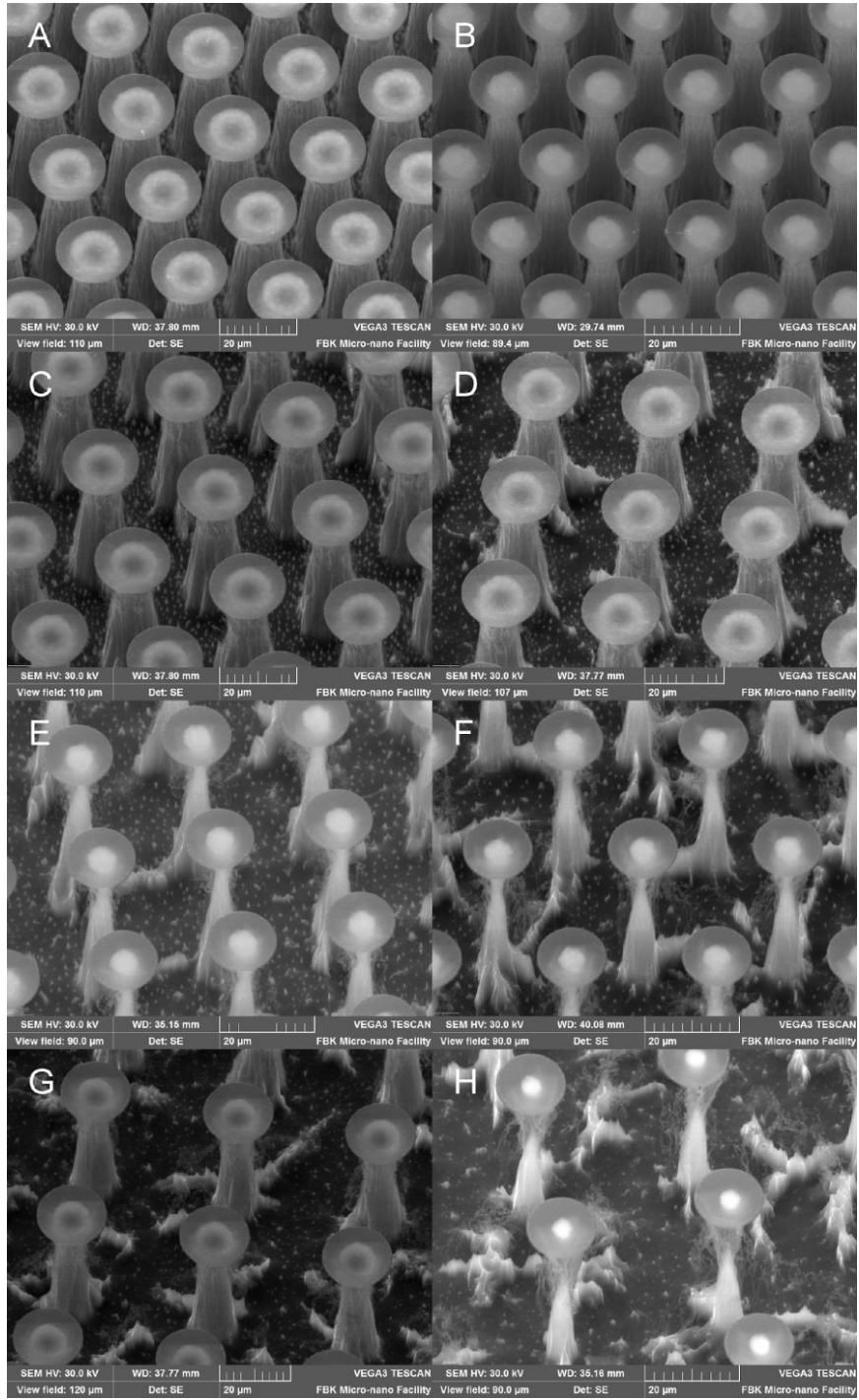

*Figure S4. A-H) SEM images of various patterns etched for 80 min correspondent to some points of the data in Figure 4.6 (green dots).*

Also, $\beta$ varies both with the etching time for each specific mask geometry (SEM images in Figure S3.b-j and data in Figure S5.A) and with $\varphi$ for a certain etching time (SEM images in Figure S4.b-h and data in Figure S5.c-d). With the same mask geometry there is a peaked trend with time: for small etching times $\beta \sim 1 \div 4°$, for intermediate times it has a peak with $\beta \sim 4 \div 8°$ and for large times (when in some cases the under etching detaches the mask) it decreases towards $\beta \sim 4°$. For etching time of 26, 45 or 55 min, $\beta$ decreases as $\varphi$ increases while for etching times of 65 or 80 min the $\beta$ trend is peaked. Not only the tapering but also the etch rate varies with the etching time (Figure S5.b): the trend is similar to the one of $\beta$ but the intermediate range is more flat with a mean etch rate of 680 ± 80 nm/min.

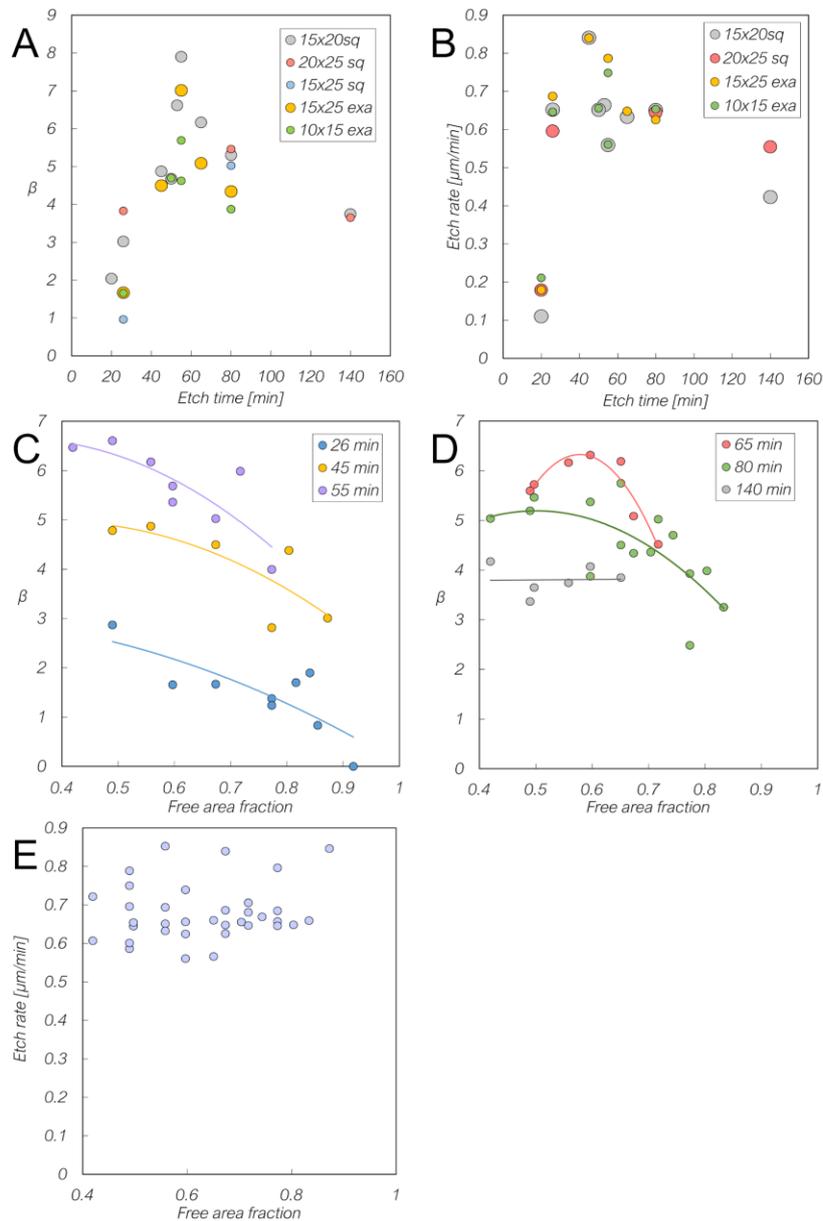

*Figure S5. A) Tapering $\beta$ vs etching time for five surfaces and B) etch rate vs time for four surfaces. C-D) tapering $\beta$ vs free area fraction after 26, 45, 55, 65, 80 and 140 min of etching (solid lines are polynomial fits) and E) etch rate vs free area fraction for the splits of Mask_1.*

The characterization of the recipe provides several information for the design of microcones with different size (to study self-ejection for droplet with different size) but equal tapering: free area fraction greater than 0.7 leads to the formation of lateral and bottom ribs, undesirable for uniformity, as well as having a small $\beta$ (<4°); in the intermediate times (45÷80 min), the etch rate is independent of the free area fraction
 (Figure S5.e); for intermediate time and for small $\varphi$ tapering is on average ~7° and microcones exhibit the highest uniformity. Thus, for Mask2 we chose $\varphi$~0.55.

The next step is the removal of the hard mask and the polymer (passivation layer deriving from the polymerization of $C_4F_8$, visible in Figure S3.j and S4.h). We tried two methods:
1. HF vapours (PRIMAXX® uEtch®) to etch silicon oxide under Al disks, Al disks removal with a jet of deionized water and oxygen plasma to remove the polymer. Removal of the mask was successful but polymer etching was not effective. Indeed, the polymer remains, as clearly visible in Figure S6.a after an isotropic etching (Tegal 900) of about 2 µm depth.

2. We therefore opted for immersion in isopropanol with ultrasonic pulses followed by cleaning with a jet of deionized water; for some patterns the ultrasounds break the disks (Figure S6.b). Then, a dip in Al etch solution and in a Piranha solution was performed to remove the hard mask (Figure S6.c). Finally, to remove the slight overhang on top, an isotropic etching (Tegal 900) was employed to lower the pillars of about 2 µm (Figure S6.d).

The microcones of Mask2 are shown in Figure 1.c-f and their parameters are in Table S2. The tapering is the same for smaller cones (splits similar to Mask1) while it is higher for larger ones. However, the taper $\beta$ is that of the medium-low part of the cones, as towards the apex they are straight. Since self-ejection involves the tapering of the medium-high part (where the drop is positioned before the jump) we measured a medium tapering angle $\underline{\beta}$ (Table S2) for that zone as shown in Figure S8 from lateral images captured with the camera. $\underline{\beta}$ is on average similar for the four surfaces tested in condensation experiments.

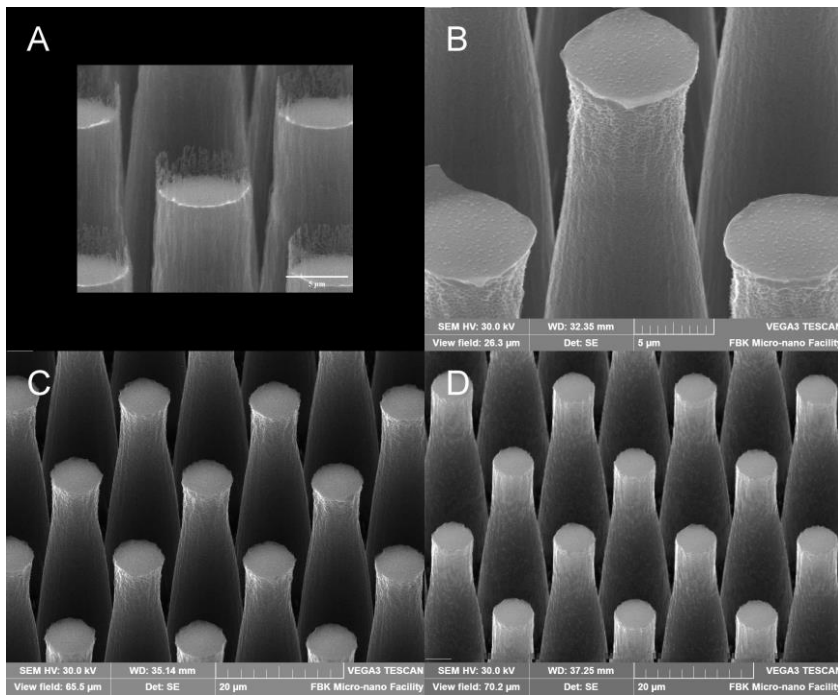

Figure S6. *a*) Detail of the residual polymer of Method 1. Method_2: *b*) pillar after the ultrasounds in isopropanol. The overhanging part of the disks is broken. *c*) Microcones after wet etch of the mask and *d*) after the isotropic etching.

Table S2. Parameters of the Mask2 and of the microcones tested in condensation conditions.

| Surface name | $D_{mask}$ [µm] | Pitch $p$ [µm] | $\beta$ [°] | $\underline{\beta}$ [°] | Head diameter $d_h$ [µm] | Height [µm] |
|---|---|---|---|---|---|---|
| 10x13 | 10 | 13 | 6.6 | 5.8 ± 0.7 | 5 | 23.3 |
| 15x20 | 15 | 20 | 6.6 | 5.7 ± 1 | 7 | 34.9 |
| 30x40 | 30 | 40 | 10 | 5.8 ± 0.6 | 12.5 | 64.3 |
| 60x80 | 60 | 80 | 8.8 | 5.5 ± 0.2 | 31.5 | 103.7 |

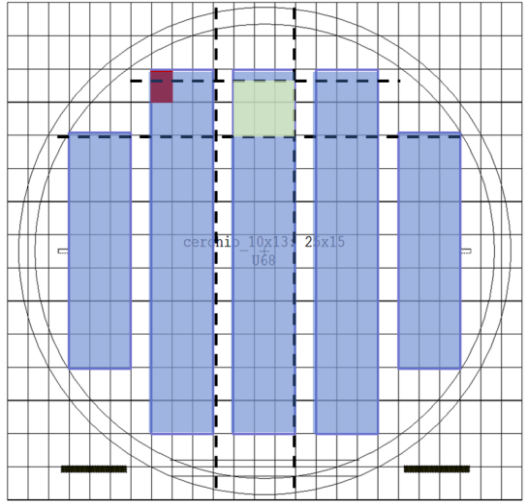

*Figure S7. Lithography of the Mask_2 with the stepper technique. The pattern of circles is exposed on the blue areas one red rectangle at a time. The dotted lines are an example of cleavage of the wafer, the green area indicating a cleaved sample (~2 cm x 2cm ).*

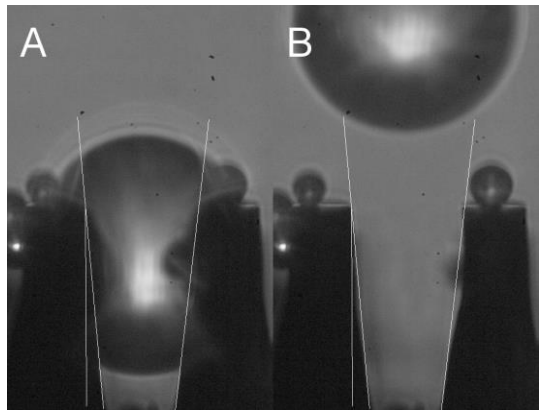

*Figure S8. Measurement of $\underline{\beta}$ from side-view frames. $\underline{\beta}$ is an average tapering of the self-ejection zone.*

## S5.2 Contact angles and NanoAl

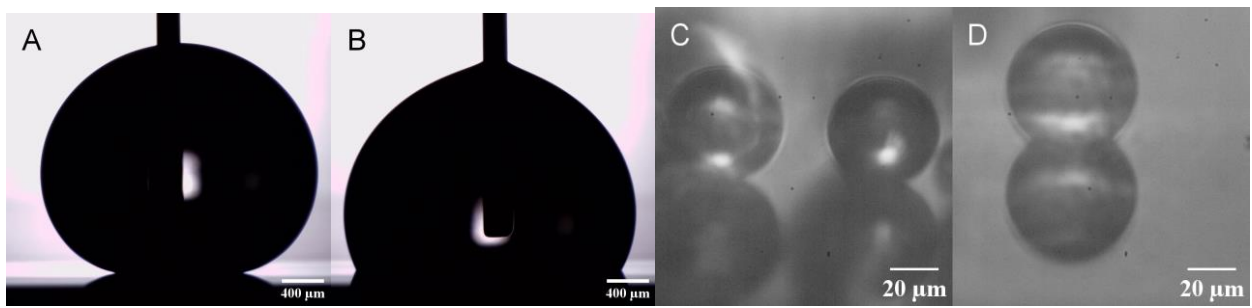

*Figure S9. A) and B) are frames captured during advancement and recession of the contact line from which we measure $\vartheta_a$ and $\vartheta_r$, respectively, with the macro-droplets method. C) and D) come from the micro-droplet method.*

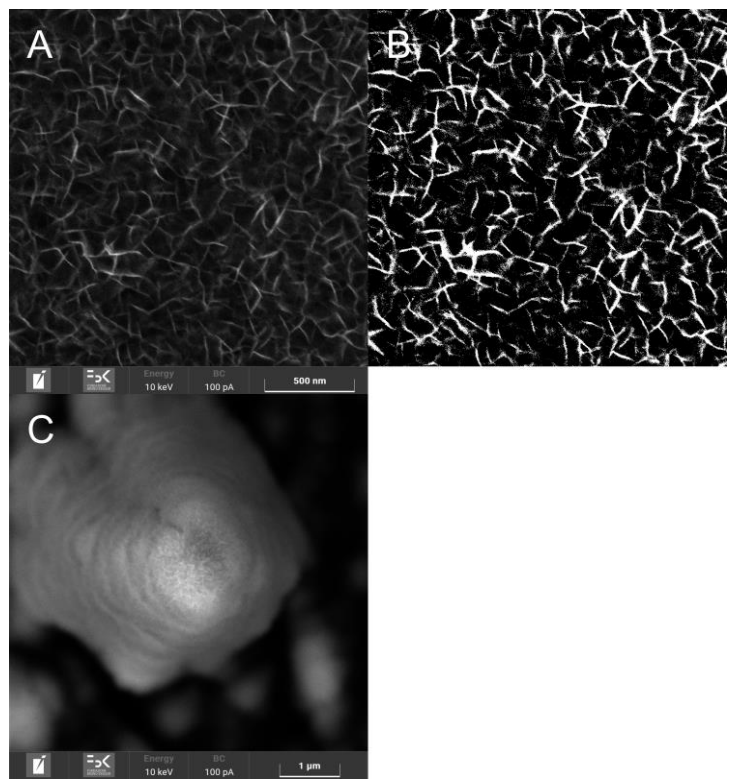

*Figure S10. SEM images of NanoAl: A) NanoAl_5min HWT on a piece of flat silicon covered by Al. B) Image analysis of A). C) An example of microcone covered by NanoAl.*